\documentclass[letterpaper,11pt,reqno]{article}
\usepackage{paper}
\usepackage{float}
\usepackage{array}
\usepackage{tabularx}
\usepackage{algorithm}
\usepackage{algpseudocode}

\hypersetup{pdftitle={Sequential Compliance Decisions of Firms on Cross-Border Data Flows: An Institutionally Anchored Decision Support System}}

\begin{document}

\title{Sequential Compliance Decisions of Firms on Cross-Border Data Flows: An Institutionally Anchored Decision Support System}

\author{Yuepeng Zhou, Dongchi Xing, Li Xiong\textsuperscript{*}
%
\thanks{Yuepeng Zhou, Dongchi Xing, Li Xiong: School of Management, Shanghai University, Shanghai 200444, China. \textsuperscript{*}Corresponding author: Li Xiong (xiongli7@126.com). This work was supported by the National Social Science Fund Postdoctoral Research Funding Project of China (No.~24FGLB114).}}

\date{July 2026}


\begin{titlepage}
\maketitle

The economic value of data arises from its flow across organizations and national borders. Yet increasingly stringent data governance regimes are turning cross-border transfer into an institutionally constrained sequential decision, in which firms repeatedly weigh compliance costs against the value of data flows. From the perspective of a data-exporting firm, this paper develops an institutionally anchored decision support system. It converts regulatory rules into a computable minimal compliance mapping and models the firm's weekly decisions as a finite-horizon Markov decision process (MDP), with compliance represented as a hard constraint rather than a penalty term. The resulting problem is solved using masked deep reinforcement learning, while counterfactual path advantages provide interpretable signals to support the firm's cross-border data flow decisions. Experiments show that the policies learned within the system outperform the baselines considered and deliver interpretable, auditable decision support. Local processing concentrates in states where the business value of small lawful transfers does not cover their compliance costs, and the localization boundary shifts systematically as the regime tightens. Credential acquisition is front-loaded within the compliance year, and shallow decision trees reproduce the policy's decisions with high fidelity. Treating the persistent-friction weight as a continuous representation of regulatory strictness further reveals an absorb-then-adjust pattern, in which expected rewards decline before observable behavior changes, implying that assessments based only on behavioral indicators may understate the burden already borne by firms. Moreover, the system is not tied to any specific regulation and can be transferred to other jurisdictions and rule-based compliance problems.


\end{titlepage}

\section{Introduction}\label{s:intro}

Data has become a nonrival factor of production whose economic value is often realized through reuse and flows across teams, firms, and national borders \citep{Goldfarb19,Jones20}. Yet these value-creating flows of data have become a persistent focus of data governance regimes across jurisdictions. Organized around data sovereignty, data localization, and personal information protection, overlapping and differentiated regulatory regimes have transformed data transmission from a largely technical matter into an institutionally constrained firm decision. Cross-country regulatory differences, in turn, translate directly into firm-level compliance costs \citep{Rong25}. OECD/WTO estimates indicate that, if all economies fully restricted data flows and moved to a completely fragmented data regime, global GDP would fall by about 4.5\% and exports by about 8.5\%, whereas open data flows with safeguards would increase output \citep{OECDWTO25}. Cross-border data governance is therefore not an ancillary compliance issue but a central concern in the functioning of the digital economy. Facing such regimes, firms must balance the compliance costs of lawful transfer against the business value generated by cross-border data transfer. This trade-off also has an investment character: compliance outlays are incurred up front, whereas trust-based returns materialize only over the long term, creating a spend-first, benefit-later pattern \citep{Chisam26}. Existing research, however, mostly treats this trade-off as a one-off strategic choice, even though firms confront it repeatedly in ongoing operations.

China's Provisions on Promoting and Regulating the Cross-Border Flow of Data \citep{CAC24} provide a representative institutional setting for this broader governance trend. Within the framework established by the Personal Information Protection Law \citep{NPC21}, the Provisions set escalating compliance requirements for personal information export according to the data type, the scale of data subjects, and the firm's cumulative volume of exported personal information. The applicable mechanisms include exemption under statutory scenarios; conclusion of a standard contract for the outbound cross-border transfer of personal information (hereafter the standard contract) or certification for personal information protection (hereafter protection certification); and a government-led security assessment \citep{CAC22,CAC23,CAC24}. At the same time, the 2024 revision reduced part of the burden through scenario-based exemptions and free trade zone (FTZ) negative lists, which vary across regions \citep{CAC24,Ma25}. For firms subject to this regime, compliance is therefore not a one-off legal judgment, but an institutional constraint embedded in operations.

From the perspective of the choices a manager actually faces each week, the regime creates a dynamic decision problem with three coupled dimensions: how much data to transfer in the current week, whether to transfer the data across the border or turn to local processing, and whether and when to invest in a compliance credential that reduces subsequent compliance costs. Two features make this problem sequential rather than a series of independent one-period decisions. First, the firm cannot optimize its export mechanism solely on economic grounds because regulation imposes a minimum level of compliance stringency. The firm can adjust the transfer volume, choose between cross-border transfer and local processing, and determine the timing of credential investment, but these decisions must remain above the regulatory floor. Second, the firm's own decision history reshapes its future feasible actions and cost structure. Cumulative personal-information export volumes can raise the minimum compliance stringency required in later weeks; sustained export activity accumulates into regulatory friction; and an acquired credential may be invalidated by a material business change. A choice that is inexpensive in the current week can therefore raise the compliance requirements and operating costs faced in subsequent weeks. Firm-side cross-border data transfer is thus a dynamic decision problem under endogenous, history-dependent constraints, a structure that static compliance checklists and single-period heuristics cannot capture.

Despite its practical significance, the dynamic firm-side decision problem has received limited attention in existing research. On the one hand, RegTech and compliance analytics mainly cast compliance as a prediction or classification task, typically from the regulator's perspective. Regulatory signals are treated as labels to be predicted ex post rather than as institutional constraints that delimit the firm's feasible choices ex ante \citep{Siering22}. On the other hand, international business and legal studies that adopt the firm's perspective document the costs of cross-border data compliance and their cross-jurisdictional differences, but they typically characterize firm responses as fixed localization or transfer strategies. They rarely examine how firms make sequential decisions under a shifting feasibility boundary \citep{Ma25,Rong25}. Decision-support research has shown that deep reinforcement learning can transform operational problems with this sequential character into actionable and interpretable policies \citep{Zhou25}. In firm-side cross-border data compliance, however, legal rules directly delimit the feasible action set, these constraints evolve with the firm's own history, and a key decision concerns when to invest in a credential that may subsequently be invalidated. These features fall outside the environmental settings typically considered in existing decision support studies. Accordingly, this study addresses three research questions: when a firm should turn to local processing rather than transfer at the minimum compliance stringency; whether and when it should invest in a costly credential that may be invalidated; and how these decision boundaries shift as the institutional environment tightens. Addressing these questions requires more than applying an existing learning algorithm. The regulatory rules must first be converted into a computable decision environment so that institutional constraints enter the firm's sequential decision process as boundaries on feasible actions.

To this end, this paper proposes an institutionally anchored decision support system. Specifically, the data-export routing logic of the Provisions on Promoting and Regulating the Cross-Border Flow of Data is abstracted into a computable minimal compliance mapping. Given the firm's weekly state, the mapping returns the minimum compliance stringency required by the regime. On this basis, the firm's decisions over cross-border transfer, local processing, and credential investment are then formulated as a finite-horizon Markov decision process (MDP). Within this system, compliance requirements enter the decision environment as legal action sets, forming hard constraints on the feasible region rather than penalty terms in the reward function \citep{Wachi24}. On the solution side, the paper introduces counterfactual path-advantage augmentation (CPAA), which predicts the long-run value of four strategic paths offline and incorporates these signals into the policy observation. Masked reinforcement learning then learns a policy within the state-dependent legal action set \citep{Stolz24}. The resulting path-value signals support policy computation while also providing managers with a readable basis for understanding the policy's preferences.

Whereas existing compliance analytics treat firm behavior as an ex-post classification target, this paper derives a firm's state-dependent legal action set directly from statutory cross-border data rules and couples it with interpretable path-value signals, thereby recasting firm-side compliance as a forward-looking sequential decision problem. The contributions of this paper are threefold. (1) The firm-side cross-border data compliance problem is defined as an institution-constrained sequential decision problem whose feasibility boundary is delimited by regulatory rules and evolves endogenously with the firm's transfer history. This formulation captures a decision structure that static compliance checklists and single-period heuristics cannot represent. (2) An institutionally anchored decision support system is proposed to solve this problem. The minimal compliance mapping converts statutory boundaries into state-dependent legal action sets, so that compliance constraints enter the decision process as hard boundaries. CPAA characterizes the long-run value of the strategic paths, and masked reinforcement learning learns a compliance policy within the legal action set. (3) The learned transfer policy is translated into interpretable decision support for managers. CPAA provides readable path-value signals, and shallow decision trees distill actionable rules. Together, these analyses address the sequential decision questions of the study: local processing concentrates in states where the business value of small lawful transfers cannot cover the compliance costs, and this boundary shifts as the institutional environment tightens; credential investment tends to be front-loaded within the compliance year.

The remainder of this paper is organized as follows. Section~\ref{s:litreview} reviews the literature on the value and institutional constraints of cross-border data flows, static characterizations of firm compliance, and constrained sequential decision-making with interpretable decision support. Section~\ref{s:model} develops the institutionally anchored decision environment, including the minimal compliance mapping, the Markov decision process, credential-investment dynamics, and the legal action set. Section~\ref{s:method} presents the solution method, comprising the CPAA counterfactual path-prediction module and the masked reinforcement learning agents based on augmented observations. Section~\ref{s:experiments} reports the numerical experiments and decision support analyses. Section~\ref{s:ccl} concludes with the main findings, the study's limitations, and directions for future research.

\section{Literature review}\label{s:litreview}

\subsection{Value and institutional constraints of cross-border data flows}

Data is a nonrival factor of production whose economic value lies not in static possession but in reuse across parties and purposes. The same dataset can support multiple uses simultaneously without physical depletion. Pooling data can also yield scale benefits beyond what an individual firm could achieve, and these benefits increase as the scope of reuse expands \citep{Jones20,Farboodi23}. Much of the value of data is therefore realized through flows across teams, firms, and markets. Restrictions on where data may move can suppress this value creation and impose opportunity costs on the firms that hold and use data \citep{Goldfarb19}. Data flows also generate externalities that complicate the allocation of the resulting value among parties \citep{Acemoglu22}, but this does not alter their central role in generating business value. For firms operating across national borders, cross-border data transfer is not merely a technical operation or legal formality but an important source of realizing business value.

These value-creating flows have also become the primary object of data governance regimes. Jurisdictions have developed overlapping yet differentiated regimes around data sovereignty, data localization, and personal information protection \citep{Aaronson18,Mattoo18}. Their differences do not offset one another; rather, they compound the compliance costs that firms face when transferring data across borders \citep{Rong25}. Information systems research further shows that institutional arrangements governing data sharing and data siloing can alter innovation incentives and social welfare in platform ecosystems \citep{Kramer25}. At the aggregate level, the costs of a fully fragmented data regime for global output and exports have been quantified \citep{OECDWTO25}. Cross-border data governance is thus not a secondary compliance concern but an institutional issue with direct implications for the efficiency of the digital economy.

As institutional constraints tighten, local processing may appear to offer firms an alternative around the compliance burden of transfer, but the option is itself costly. Research on data localization shows that requirements to keep data and related activities within national borders can raise firms' data-management and operating costs and reduce operational flexibility, while the broader economic case for localization remains contested. In repeated discrete choice experiments covering seven populous countries, \citet{Prince25} find that individuals generally show no significant preference for having their own data stored in particular locations, except for limited additional valuation of local storage for privacy-sensitive data such as financial and biometric information. This evidence suggests that local storage alone is unlikely to generate substantial additional demand-side revenue. From the firm's perspective, local processing therefore involves giving up at least part of the business value of cross-border transfer in exchange for lower exposure to transfer-related compliance risk \citep{OECD23}. Local processing is not a costless safe option but a potentially rational outside option in particular states.

Taken together, these studies establish the value basis of data flows, document the macro-level cost of institutional frictions, and clarify the role of local processing as an alternative path. However, they remain largely at the level of institutions, policies, and aggregate outcomes. They do not explain how an individual firm should make period-by-period choices under a given regulatory regime. In other words, existing research explains why cross-border data flows matter but says little about how a firm that cannot change the institutional boundary should decide sequentially among cross-border transfer, local processing, and compliance investment.

\subsection{Static characterizations of firm compliance}

Although the firm-side sequential decision problem is of practical significance, existing compliance research has not adequately covered it. The literature falls broadly into two strands. One takes the regulator's perspective and treats compliance as a prediction, classification, or verification task. The other takes the firm's perspective and focuses on the costs that cross-border data regimes impose on firms. Their limitations lie along two dimensions, research perspective and the time structure of decisions, and neither strand characterizes how firms decide repeatedly within institutional boundaries.

The first strand covers compliance analytics and regulatory technology (RegTech), which commonly frame compliance as the identification or evaluation of observed behaviors, transactions, or processes. In financial and accounting fraud detection, violations are typically modeled as low-incidence labels predicted by data mining or machine learning \citep{Ngai11,Abbasi12,Bao20}. In RegTech and regulatory reporting, related techniques support automated monitoring, reporting, and enforcement \citep{Siering22}. Business-process compliance research, in turn, formalizes regulatory requirements as rules for checking whether existing processes violate constraints \citep{Hashmi18}. Despite their methodological differences, these studies focus on behavior or processes that have already occurred and ask whether they satisfy regulatory criteria. They therefore support ex-post detection or supervision, with the firm positioned mainly as the object of scrutiny rather than as a decision-maker making forward-looking choices under institutional constraints.

The second strand turns to the firm's perspective and examines the actual impact of data regulation on firms. One line of empirical research quantifies the effects of privacy and data regulation on firms. For example, studies report that the General Data Protection Regulation (GDPR) reduced venture investment through compliance costs and intensified market concentration toward large firms \citep{Jia21,Johnson23}. International business and legal studies further characterize how cross-country differences in data regimes translate into firm burdens \citep{Rong25,Ma25}. They also note the investment character of compliance, whereby outlays are incurred up front while returns materialize only over the longer term \citep{Chisam26}. These literatures establish that firms bear substantial and heterogeneous compliance costs. But they generally characterize firm responses as static choices rather than as repeated adjustments to changing external conditions and the firm's own decision history.

In sum, existing firm-compliance research falls short in two respects. RegTech and compliance analytics have supervisory value, but they primarily adopt the perspective of regulators or auditors and emphasize ex-post evaluation. Firm-side compliance research recognizes the burdens imposed on firms but generally characterizes their responses as phased static choices. Neither addresses the problem here, in which institutional rules determine the firm's compliance requirements, the stringency of those requirements evolves endogenously with the firm's own history, and the firm must repeatedly choose among cross-border transfer, local processing, and credential investment. Characterizing such a problem requires a decision environment with an evolving institutional boundary and a method capable of supporting sequential decisions within that boundary.

\subsection{Constrained sequential decision support}

The transfer, localization, and credential-investment decisions that firms make repeatedly under institutional constraints can be formulated as a Markov decision process. When the state space and transition dynamics make exact dynamic programming infeasible, deep reinforcement learning (DRL) provides an important solution approach for sequential decision support problems. Prior studies show that, in operational settings combining repeated decisions, stochastic exogenous inputs, and long-run payoff trade-offs, DRL can learn executable policies without requiring a closed-form solution. Applications include inventory control, dual sourcing, multi-echelon networks, and revenue management \citep{Gijsbrechts22,Chen23}. Most of this research takes the platform or operations planner as the decision-maker. By contrast, \citet{Zhou25} adopt the perspective of an individual courier operating under platform rules, formulate repositioning and order-acceptance decisions as a Markov decision process, and use DRL to generate actionable and interpretable decision support. Most studies learn policies end to end, whereas \citet{Harsha25} embed an optimization subproblem at each decision step, integrating integer programming into policy iteration to guide replenishment decisions. In these settings, however, the constraints are specified as operational structures. They are not derived from statutory text, nor do they evolve as a consequence of the decision-maker's own history.

Firm-side data compliance introduces an additional feature that distinguishes it from conventional operational decision problems: legal rules delimit the set of admissible actions. A transfer below the minimum compliance stringency is not an available option that merely carries a higher cost; it is legally infeasible. Research on constrained reinforcement learning studies how to optimize reward without violating such hard constraints, either by expressing feasibility through auxiliary cost constraints or by excluding infeasible actions directly from the decision process \citep{Wachi24}. Among these approaches, action masking removes illegal actions from the policy's choice set instead of penalizing them in the reward, which aligns most directly with state-dependent hard legal boundaries. Its learning-efficiency advantage increases as the share of infeasible actions grows, and it has been extended from discrete to continuous action spaces \citep{Huang22,Stolz24}. Existing literature, however, typically treats the environment and constraints as given. The derivation of state-dependent legal action sets from concrete statutory rules has received limited attention, as has the question of how the decision-maker's own history endogenously changes the constraints faced in subsequent periods.

Even with the solution paradigm in place, a model cannot serve as an effective decision support tool if managers cannot understand the basis of its decisions. The requirement is especially salient in regulated settings such as data-export compliance, where a compliance manager needs not only recommendations but also an explanation of why cross-border transfer, local processing, or credential investment is chosen in a particular state. Interpretability therefore becomes an integral part of the decision support problem. \citet{Rudin19} argues that high-stakes decisions should rely on interpretable models rather than post-hoc rationalizations of black-box predictions. In reinforcement learning, policy interpretability is commonly pursued along two broad routes. One distills a trained policy into a compact and verifiable proxy, such as a decision tree or concept-based representation \citep{BastaniO18}. The other embeds interpretability directly into the learning process, as summarized by \citet{Glanois24}. Management research has likewise used reinforcement learning to extract readable rules that can improve human sequential decision-making \citep{BastaniH26}.

\subsection{Research gap and research questions}

Taken together, the preceding literature leaves three related issues unresolved. Data economics and governance research explains the value of data flows, the macro-level cost of institutional frictions and the economic trade-offs associated with local processing, but it remains at the level of policies and aggregate outcomes. Firm-side compliance research attends to firm-level costs and cross-country differences, but it either emphasizes ex-post identification from the regulator's perspective or characterizes firm responses as static choices. Constrained sequential decision-making and interpretable reinforcement learning provide tools such as action masking and policy distillation, but they usually treat the decision environment and constraints as exogenously given. How concrete regulatory rules can be converted into state-dependent legal action sets, how those action sets evolve with the firm's own history, and how the resulting policy can be delivered as interpretable and auditable decision support remain underdeveloped.

The gap addressed in this study is therefore an institutionally anchored firm-side sequential decision problem in which the feasibility boundary is generated by regulatory rules, evolves endogenously with the firm's decision history, and must ultimately be translated into interpretable and auditable decision support. Accordingly, this study asks three research questions: when a firm should turn to local processing rather than transfer at the minimum compliance stringency; whether and when a firm should invest in a costly credential that may be invalidated; and how these decision boundaries shift as the institutional environment tightens. The first two questions characterize the localization and credential-investment dimensions of the learned policy, whereas the third examines how changes in the institutional environment shift the firm's behavioral boundaries. Table~\ref{t:table1} positions this study relative to the most closely related literature along four dimensions: research perspective, modeling approach, solution methodology, and model outcome.

\begin{table}[H]
\caption{Positioning of this work relative to the most related prior studies}
\begin{tabularx}{\textwidth}{@{}>{\hsize=0.8\hsize\raggedright\arraybackslash}X>{\hsize=0.9\hsize\raggedright\arraybackslash}X>{\hsize=1.0\hsize\raggedright\arraybackslash}X>{\hsize=1.05\hsize\raggedright\arraybackslash}X>{\hsize=1.25\hsize\raggedright\arraybackslash}X@{}}
\toprule
Reference & Perspective & Modeling approach & Solution methodology & Model outcome\\
\midrule
\citet{Siering22} & Regulator / auditor & Supervised classification & RegTech monitoring & Non-compliant cases flagged ex post\\
\citet{Chisam26} & Firm, cross-jurisdiction & Cross-country empirical study & Econometric analysis & Privacy investment trade-off as a static posture\\
\citet{Gijsbrechts22} & Operations planner & Inventory MDP & Deep RL & Near-optimal operational policy\\
\citet{Harsha25} & Operations planner & Inventory MDP with integer programming & Deep RL with in-loop optimization & Constrained replenishment policy\\
\citet{Zhou25} & Single agent (courier) & Operational MDP & Deep RL & Actionable, interpretable decision support\\
\citet{BastaniO18} & Policy verifier & Pretrained RL policy & Decision-tree distillation & Verifiable interpretable proxy\\
\textbf{This work} & \textbf{Firm (data exporter)} & \textbf{Institutionally anchored MDP} & \textbf{Masked deep RL (D3QN+CPAA)} & \textbf{Interpretable, auditable decision support}\\
\bottomrule
\end{tabularx}
\label{t:table1}\end{table}

\section{Problem and decision environment}\label{s:model}

\subsection{Environment overview and notation}

\begin{figure}[H]
\includegraphics[width=\textwidth,page=1]{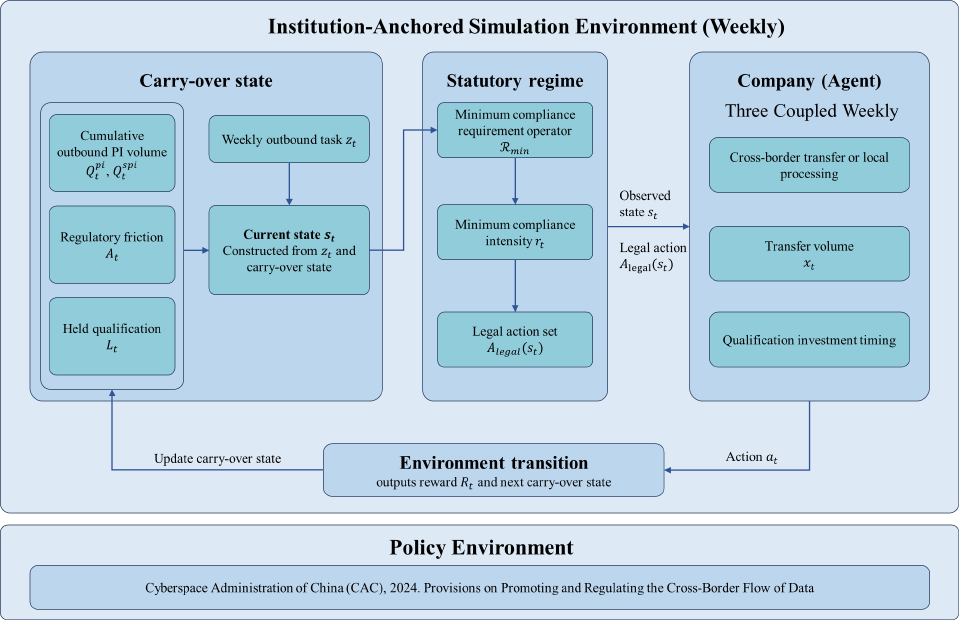}
\caption{The simulation environment from the perspective of a data-exporting firm}
\label{f:fig1}\end{figure}

This paper develops a stochastic simulation environment that evolves in weekly increments to characterize a firm's operations under the data-export regime established by the Provisions on Promoting and Regulating the Cross-Border Flow of Data \citep{CAC24}. The environment consists of three parts: weekly data-export demand, a minimal compliance mapping that converts statutory provisions into compliance requirements, and an evolving compliance state that links the firm's decision history to its week-by-week economic payoffs. Each week, the environment generates one export task carrying the data type, business type, destination group, statutory exemption scenario, and demand volume. Upon receiving the exogenous export task, the firm chooses between cross-border transfer and local processing in light of the current compliance requirement and its historical state. It also determines when, if at all, to invest in a compliance credential within the compliance year. Fig.~\ref{f:fig1} presents the simulation environment from the firm's perspective.

The firm's state includes the data type, firm attributes, and cumulative personal-information export volumes, among other components. The minimal compliance mapping determines the minimum compliance stringency required in the current week. Cumulative personal-information export volumes evolve with the firm's transfer history and can raise the minimum compliance stringency required in later weeks, while regulatory friction and compliance credentials link current decisions to future costs. The problem is therefore an intertemporally coupled sequential decision problem rather than a series of independent single-period decisions. Given the minimum compliance stringency in each week, the firm faces three coupled decisions: whether to transfer data across the border or process locally, how much data to transfer, and when to invest in a compliance credential. Table~\ref{t:table2} summarizes the notation used in the model and solution method.

\begin{table}[H]
\caption{Modeling notation and definitions}
\begin{tabular*}{\textwidth}[]{@{\extracolsep\fill}p{4.6cm}p{9.8cm}}
\toprule
Symbol & Definition\\
\midrule
\multicolumn{2}{l}{\textit{Indices and sets}}\\
$t,\ T$ & Week index and horizon, $t=0,\dots,T-1$, $T=52$\\
$\tilde{\mathcal{S}},\ \mathcal{S}$ & Base state space and observation space, $\mathcal{S}=\tilde{\mathcal{S}}\times\mathcal{G}$\\
$\mathcal{A}$ & Action space (a response path paired with a transfer volume)\\
$\mathcal{G}=\{E,M,H\}$ & The three compliance stringency tiers, $E\prec M\prec H$\\
$\mathcal{E}_{\mathrm{exempt}}$ & Set of statutory exemption scenarios\\
$\mathcal{R}_{\min}$ & Minimal compliance mapping from the base state to the week's minimum compliance stringency ($\tilde{\mathcal{S}}\to\mathcal{G}$)\\
\multicolumn{2}{l}{\textit{Exogenous weekly task $z_t=(d_t,\tau_t,k_t,e_t,q_t)$}}\\
$d_t$ & Data type, $\{\mathrm{GEN},\mathrm{PI},\mathrm{SPI},\mathrm{IMPORTANT}\}$\\
$\tau_t$ & Business type\\
$k_t$ & Destination group\\
$e_t$ & Statutory exemption or GBA special-mechanism scenario\\
$q_t$ & Weekly demand volume\\
\multicolumn{2}{l}{\textit{Decision variables (action $a_t=(m_t,x_t)$)}}\\
$m_t$ & Response path, $\{\mathrm{EXEMPT},\mathrm{SCC},\mathrm{CERT},\mathrm{SA},\mathrm{LOCAL}\}$\\
$x_t$ & Data transfer volume, $0\le x_t\le q_t$; $x_t\equiv 0$ under $\mathrm{LOCAL}$\\
\multicolumn{2}{l}{\textit{State variables}}\\
$\mathrm{CIIO}$ & Critical information infrastructure operator flag, $\{0,1\}$\\
$u_t$ & Regional policy status, $\{\mathrm{NORMAL},\mathrm{FTZ},\mathrm{GBA}\}$\\
$Q^{pi}_t,\ Q^{spi}_t$ & Within-year cumulative export volumes of non-sensitive and sensitive personal information\\
$A_t$ & Regulatory friction stock, $A_t\in[A_{\min},A_{\max}]$\\
$L_t$ & Highest credential currently held, $\{0,1,2\}$\\
$r_t$ & The firm's current minimum compliance stringency for data export, $r_t\in\mathcal{G}$ (observable)\\
$\tilde s_t,\ s_t$ & Base state and observation, $s_t=(\tilde s_t,r_t)$\\
\multicolumn{2}{l}{\textit{Parameters}}\\
$\gamma$ & Discount factor\\
$\beta$ & Curvature of the business-value function\\
$\mu$ & Shortfall rate on unmet demand\\
$\alpha$ & Friction memory\\
$\kappa_A,\ \kappa_\sigma$ & Friction-cost weights on the stock and on the per-period increment\\
$p_{\mathrm{chg}}$ & Credential invalidation risk\\
$q_{\mathrm{ref}}$ & Demand normalization scale\\
\bottomrule
\end{tabular*}
\label{t:table2}\end{table}

\subsection{The minimal compliance mapping}

Each week, the firm must determine the minimum compliance stringency required for lawful data export. To make the institutional requirement computable, the data-export provisions are formalized as a minimal compliance mapping, denoted by $\mathcal{R}_{\min}$. Given the firm's base state in the current week, the mapping returns one of three minimum compliance stringency tiers:
\begin{equation}
\mathcal{R}_{\min}:\tilde{\mathcal{S}}\to\mathcal{G},\qquad \mathcal{G}=\{E,M,H\},\qquad E\prec M\prec H,
\label{e:eq1}\end{equation}
where the tiers are ordered from the lowest to the highest stringency. Tier $E$ (exemption) permits transfer on the basis of a statutory exemption scenario alone; tier $M$ (standard) requires at least the standard contract or protection certification; whereas tier $H$ (assessment) requires the government-led security assessment \citep{NPC21,CAC22,CAC23,CAC24}. The mapping therefore yields an unambiguous minimum requirement against which an export mechanism can be classified as exactly compliant, over-compliant, or infeasible.

The regulation sets statutory thresholds for cumulative personal-information export volumes, and crossing these thresholds escalates the required compliance stringency \citep{CAC24}. Let $Q^{pi}_t$ and $Q^{spi}_t$ denote the firm's cumulative export volumes of non-sensitive and sensitive personal information within the compliance year (reset at calendar-year boundaries). The thresholds affect the stringency returned by $\mathcal{R}_{\min}$ but do not constrain the single-period transfer $x_t$. After a threshold is crossed, the firm may continue to transfer data, but only through a mechanism of higher stringency. Accordingly, $Q^{pi}_t$ and $Q^{spi}_t$ are endogenous state variables accumulated through the firm's own export history rather than exogenous quotas. The output of $\mathcal{R}_{\min}$ therefore depends on the firm's prior transfer decisions.

The minimum compliance stringency is evaluated through priority-ordered layers. The first triggered provision at the highest applicable priority determines the result, and lower-priority provisions are not evaluated once a higher-priority layer returns a conclusion. Accordingly, $\mathcal{R}_{\min}$ is composed of three layers arranged in descending priority. Let $v_1,v_2,v_3$ denote the partial conclusions returned by these layers, with $\bot$ indicating that a layer is not triggered. The three layers are defined as follows:
\begin{equation}
v_1(\tilde s)=\begin{cases} H, & d=\mathrm{IMPORTANT}\\ E, & e\in\mathcal{E}_{\mathrm{exempt}}\\ M, & e=\mathrm{GBA}\\ \bot, & \text{otherwise},\end{cases}
\label{e:eq2}\end{equation}
\begin{equation}
v_2(\tilde s)=\begin{cases} H, & \mathrm{CIIO}\wedge d\in\{\mathrm{PI},\mathrm{SPI}\}\\ E, & \mathrm{CIIO}\wedge d=\mathrm{GEN}\\ \bot, & \text{otherwise},\end{cases}
\label{e:eq3}\end{equation}
\begin{equation}
v_3(\tilde s)=\begin{cases}
H, & d=\mathrm{SPI},\ Q^{spi}_t+q_t\ge 10^{4}\\
M, & d=\mathrm{SPI},\ Q^{spi}_t+q_t<10^{4}\\
H, & d=\mathrm{PI},\ Q^{pi}_t+q_t\ge 10^{6}\\
M, & d=\mathrm{PI},\ 10^{5}\le Q^{pi}_t+q_t<10^{6}\\
E, & d=\mathrm{PI},\ Q^{pi}_t+q_t<10^{5}\\
E, & d=\mathrm{GEN},
\end{cases}
\label{e:eq4}\end{equation}
where $\mathcal{E}_{\mathrm{exempt}}$ contains the statutory exemption scenarios, including the specific business activities enumerated in the provisions, data transit, contractual and human-resources necessity, emergencies, and transfers falling outside free trade zone negative lists \citep{CAC24}. The minimum compliance stringency is given by the first non-empty conclusion in priority order:
\begin{equation}
\mathcal{R}_{\min}(\tilde s)=v_1(\tilde s)\,\triangleright\,v_2(\tilde s)\,\triangleright\,v_3(\tilde s),\qquad
a\triangleright b:=\begin{cases} a, & a\ne\bot\\ b, & a=\bot.\end{cases}
\label{e:eq5}\end{equation}

The three layers correspond to three classes of legal rules. Layer $v_1$ has the highest priority. Exports of important data always receive the highest stringency, and this requirement cannot be lowered by an exemption \citep{CAC22}. Moreover, statutory exemptions take precedence over volume thresholds; when an exemption applies, the exemption tier is returned \citep{CAC24}. The GBA branch corresponds to the special mechanism for the Guangdong--Hong Kong--Macao Greater Bay Area, which routes eligible transfers to the standard tier through a simplified standard contract. Layer $v_2$ imposes the special rules applicable to critical information infrastructure operators. Layer $v_3$ is the residual layer for transfers not resolved by the first two layers. It assigns compliance stringency according to cumulative volumes, with the relevant thresholds evaluated after including the current week's demand.

Because $\mathcal{R}_{\min}$ serves as a hard constraint in the decision environment, Eq.~\eqref{e:eq5} must return a valid conclusion for every state. Important data is fixed at $H$ in $v_1$ and never reaches $v_3$. Consequently, any state reaching $v_3$ involves non-important data, and the case structure in $v_3$ covers all such data types. The operator $\triangleright$ therefore always returns a non-empty conclusion. $\mathcal{R}_{\min}$ hence never produces undefined or conflicting compliance requirements, and its conclusions strictly follow statutory priority. The resulting minimum compliance stringency, $r_t=\mathcal{R}_{\min}(\tilde s_t)$, plays two roles in the Markov decision process. First, it defines the stringency floor of the legal action set. Export mechanisms whose strength falls below $r_t$ are excluded from the feasible actions, thereby converting statutory requirements into state-dependent legal constraints that evolve with the firm's transfer history. Second, it enters the policy observation as an explicit institutional signal, allowing the manager to observe the current compliance requirement directly.

\subsection{The sequential decision environment}

The firm's weekly compliance problem is modeled as a finite-horizon discounted Markov decision process whose legal action set is delimited by the minimal compliance mapping $\mathcal{R}_{\min}$. Each week, the firm observes its state, chooses a response path and a transfer volume, and receives a reward that balances business value, mechanism costs, and regulatory friction. The current decision then changes the cumulative personal-information export volumes, regulatory friction, and credential holdings, thereby affecting the feasible actions and costs of subsequent weeks. The decision epochs, states, actions, transitions, and rewards are described below.

Decisions are made weekly, and each episode represents one compliance year of $T=52$ weeks. Cumulative personal-information export volumes are reset at the beginning of each episode, consistent with the calendar-year basis of the statutory thresholds. At the start of week $t$, the firm receives an exogenous task bundle
\begin{equation}
z_t=(d_t,\tau_t,k_t,e_t,q_t),
\label{e:eq6}\end{equation}
comprising the data type $d_t\in\{\mathrm{GEN},\mathrm{PI},\mathrm{SPI},\mathrm{IMPORTANT}\}$, business type $\tau_t\in\{\mathrm{CONTRACT},\allowbreak\mathrm{HR},\allowbreak\mathrm{ANALYTICS},\allowbreak\mathrm{RISK}\}$, destination group $k_t$, statutory exemption scenario $e_t$, and demand volume $q_t$. The business type $\tau_t$ links task generation to value assessment. Different business types induce different conditional distributions of the data type $d_t$ and exemption scenario $e_t$, while $\tau_t$ also determines the business value of the transfer in the reward function. Using the same variable to govern task generation and value assessment keeps the task distribution consistent with the associated business value and facilitates the later analysis of heterogeneity across business types.

The base state contains the firm's intrinsic attributes, the current task, within-year cumulative personal-information export volumes, regulatory friction, credential level, and week index:
\begin{equation}
\tilde s_t=\bigl(\mathrm{CIIO},\,u_t,\,d_t,\,\tau_t,\,k_t,\,e_t,\,q_t,\,Q^{pi}_t,\,Q^{spi}_t,\,A_t,\,L_t,\,t\bigr).
\label{e:eq7}\end{equation}
On top of the base state, the environment generates the minimum compliance stringency $r_t=\mathcal{R}_{\min}(\tilde s_t)$ through the minimal compliance mapping, and includes this institutional signal as an additional component of the policy observation:
\begin{equation}
s_t=(\tilde s_t,\,r_t)\in\mathcal{S}.
\label{e:eq8}\end{equation}

Because $r_t$ is fully determined by $\tilde s_t$, including it in the observation does not add information to the underlying decision problem or change the attainable optimal value. Instead, it exposes a derived institutional signal that would otherwise have to be inferred from the remaining state variables. Under finite samples and function approximation, explicitly providing the minimum compliance stringency is intended to reduce the burden of representation learning, facilitate credit assignment, and support policy convergence, in the same spirit as potential-based reward shaping. During both training and evaluation, $r_t$ is recomputed from the current week's $Q^{pi}_t,Q^{spi}_t$ and remains consistent with the true state.

Actions consist of a response path and a transfer volume:
\begin{equation}
a_t=(m_t,x_t),\qquad m_t\in\{\mathrm{EXEMPT},\mathrm{SCC},\mathrm{CERT},\mathrm{SA},\mathrm{LOCAL}\}.
\label{e:eq9}\end{equation}
The first four response paths correspond to statutory export mechanisms of increasing strength. $\mathrm{LOCAL}$ denotes local processing, that is, processing within the border without cross-border transfer. For the export mechanisms, the firm chooses $0\le x_t\le q_t$; under $\mathrm{LOCAL}$ the volume is fixed at $x_t\equiv 0$. The action therefore carries three decision dimensions of the problem: the transfer volume ($x_t$), the choice between cross-border transfer and local processing, and whether to invest in a higher-level credential. Denote the mechanism strength as $\ell(\cdot)$, with $\ell(\mathrm{EXEMPT})=0$, $\ell(\mathrm{SCC})=\ell(\mathrm{CERT})=1$, and $\ell(\mathrm{SA})=2$. Local processing does not involve any export mechanism, so $\ell(\mathrm{LOCAL})=0$. Let $\ell_{\mathrm{req}}(r_t)$ denote the minimum mechanism strength required by tier $r_t$. $\mathcal{R}_{\min}$ determines the state-dependent legal action set:
\begin{equation}
\mathcal{A}_{\mathrm{legal}}(s_t)=\{\mathrm{LOCAL}\}\cup\{m\in\{\mathrm{EXEMPT},\mathrm{SCC},\mathrm{CERT},\mathrm{SA}\}:\ \ell(m)\ge\ell_{\mathrm{req}}(r_t)\}.
\label{e:eq10}\end{equation}

The firm may choose a mechanism stronger than the minimum requirement but not a weaker one, while local processing is always available. A transfer through a mechanism below the statutory minimum strength is not a suboptimal action that should merely be discouraged through a penalty; it is prohibited by the institutional rules. This regulatory floor is therefore enforced as a hard constraint. At every decision step, a mask excludes infeasible mechanisms and confines the policy to $\mathcal{A}_{\mathrm{legal}}(s_t)$, while the reward function contains no violation penalty, consistent with the logic of constrained Markov decision processes. Behavioral outcomes such as response-path shares and the localization share are used only as descriptive indicators and do not enter the optimization objective. By construction, the masked policy does not select an illegal transfer mechanism.

State transitions are generated by four conditionally independent components. Given $s_t$ and a feasible action, the environment updates the cumulative personal-information export volumes, regulatory friction, credential level, and time, and then generates a new task $z_{t+1}$.

Cumulative personal-information export volumes increase only with exports that count toward the statutory totals. Define
\begin{equation}
\chi_t=\mathbf{1}\bigl[m_t\in\{\mathrm{EXEMPT},\mathrm{SCC},\mathrm{CERT},\mathrm{SA}\}\bigr]\,\mathbf{1}\bigl[e_t\notin\mathcal{E}_{\mathrm{exempt}}\bigr]
\label{e:eq11}\end{equation}
as the indicator of whether the current week's export volume enters the cumulative count. The cumulative volumes then evolve as
\begin{equation}
Q^{pi}_{t+1}=Q^{pi}_t+x_t\,\chi_t\,\mathbf{1}[d_t=\mathrm{PI}],
\label{e:eq12}\end{equation}
\begin{equation}
Q^{spi}_{t+1}=Q^{spi}_t+x_t\,\chi_t\,\mathbf{1}[d_t=\mathrm{SPI}].
\label{e:eq13}\end{equation}

The indicator $\chi_t$ excludes two cases from the cumulative totals: local processing ($x_t=0$) and exports conducted under statutory exemptions. The latter is consistent with the priority assigned to statutory exemptions in $\mathcal{R}_{\min}$. Exports under the GBA route are not treated as statutory exemptions and therefore continue to count toward cumulative volumes.

Regulatory friction is modeled as a persistent stock that accumulates with exports and decays toward its baseline when exports subside. Define the current transfer signal as $\sigma_t=x_t/q_{\mathrm{ref}}$. The friction stock evolves according to
\begin{equation}
A_{t+1}=\operatorname{clip}\bigl(\alpha A_t+(1-\alpha)\sigma_t,\ A_{\min},\ A_{\max}\bigr),\qquad \alpha\in(0,1),
\label{e:eq14}\end{equation}
where the previous stock is retained with weight $\alpha$ and the current transfer signal enters with weight $1-\alpha$, subject to truncation within $[A_{\min},A_{\max}]$. $\sigma_t$ depends only on the transfer volume and not on the chosen mechanism, so cross-mechanism differences appear in the cost terms rather than the friction term.

The credential level $L_t$ records the highest compliance credential currently held by the firm. Holding a credential allows the corresponding mechanism to be reused at low marginal cost, although the credential may be invalidated by a material business change. Let $p_{\mathrm{chg}}$ denote the probability that such a change requires the credential to be reacquired. The credential state evolves as
\begin{equation}
L_{t+1}=\begin{cases}
0, & \text{with probability } p_{\mathrm{chg}},\\[2pt]
\max\!\bigl(L_t,\ \ell(m_t)\,\mathbf{1}[m_t\in\{\mathrm{SCC},\mathrm{CERT},\mathrm{SA}\}]\bigr), & \text{with probability } 1-p_{\mathrm{chg}}.
\end{cases}
\label{e:eq15}\end{equation}

The single-period reward is defined as the net economic payoff from the firm's weekly decision. It equals the business value realized through the transfer volume minus the corresponding export-mechanism and regulatory-friction costs:
\begin{equation}
R_t=V_t-C^{\mathrm{mech}}_t-C^{\mathrm{fric}}_t.
\label{e:eq16}\end{equation}

To capture the diminishing marginal business value of larger transfers, business value is defined as an increasing concave function of the transfer volume \citep{Goldfarb19}:
\begin{equation}
g(z)=1-e^{-\beta z},\qquad z=\frac{x_t}{q_{\mathrm{ref}}},\quad \beta>0.
\label{e:eq17}\end{equation}

The realized value is $b(\tau_t)\,g(x_t/q_{\mathrm{ref}})$, where $b(\tau_t)$ is the task-value magnitude determined by the business type. Unmet transfer demand, $q_t-x_t$, generates an opportunity loss measured on the same curve, $\mu\,b(\tau_t)\bigl[g(q_t/q_{\mathrm{ref}})-g(x_t/q_{\mathrm{ref}})\bigr]$. The resulting net business value is therefore
\begin{equation}
V_t=b(\tau_t)\Bigl[(1+\mu)\,g\!\bigl(\tfrac{x_t}{q_{\mathrm{ref}}}\bigr)-\mu\,g\!\bigl(\tfrac{q_t}{q_{\mathrm{ref}}}\bigr)\Bigr],
\label{e:eq18}\end{equation}
where the second term is constant within the week. Under $\mathrm{LOCAL}$, the firm makes no cross-border transfer, and net business value reduces to $V_t=-\mu\,b(\tau_t)\,g(q_t/q_{\mathrm{ref}})$.

Mechanism costs reflect the credential-investment structure and comprise three parts: a one-time acquisition cost $F(\cdot)$, a maintenance cost $\mathrm{maint}(\cdot)$, and a low marginal reuse cost $c_{\mathrm{var}}(\cdot)$:
\begin{equation}
C^{\mathrm{mech}}_t=F\!\bigl(\ell(m_t)\bigr)\,\mathbf{1}\{\ell(m_t)>L_t\}+\mathrm{maint}(L^{+}_t)+c_{\mathrm{var}}(m_t)\,\frac{x_t}{q_{\mathrm{ref}}},\qquad L^{+}_t=\max\!\bigl(L_t,\ell(m_t)\bigr).
\label{e:eq19}\end{equation}

The firm pays the full acquisition cost only when the strength required by the chosen mechanism exceeds the credential level it holds. A standard-contract or protection-certification credential does not offset the subsequent acquisition cost of the security assessment. Because acquisition and maintenance are constant in $x_t$, the within-period marginal cost of transfer is determined solely by $c_{\mathrm{var}}(m_t)$. The larger cost differences across mechanisms therefore arise from the acquisition and maintenance components and, consequently, from the timing of credential investment.

Friction costs enter through both the accumulated friction stock and the current increment:
\begin{equation}
C^{\mathrm{fric}}_t=\kappa_A\,A_t+\kappa_\sigma\,\frac{x_t}{q_{\mathrm{ref}}}.
\label{e:eq20}\end{equation}

A large cross-border transfer is thus penalized twice, immediately through the $\kappa_\sigma$ term and persistently in future periods through the enlarged friction stock $A_{t+1}$ (Eq.~\eqref{e:eq14}). Over the compliance year, the firm chooses a policy that maximizes expected discounted reward:
\begin{equation}
\max_{\pi}\ \mathbb{E}\Bigl[\textstyle\sum_{t=0}^{T-1}\gamma^{t}R_t\ \Big|\ \pi\Bigr],\qquad \gamma\in(0,1),
\label{e:eq21}\end{equation}
where the policy $\pi$ maps observations to legal actions.

\section{Solution method}\label{s:method}

\begin{figure}[H]
\includegraphics[width=\textwidth,page=2]{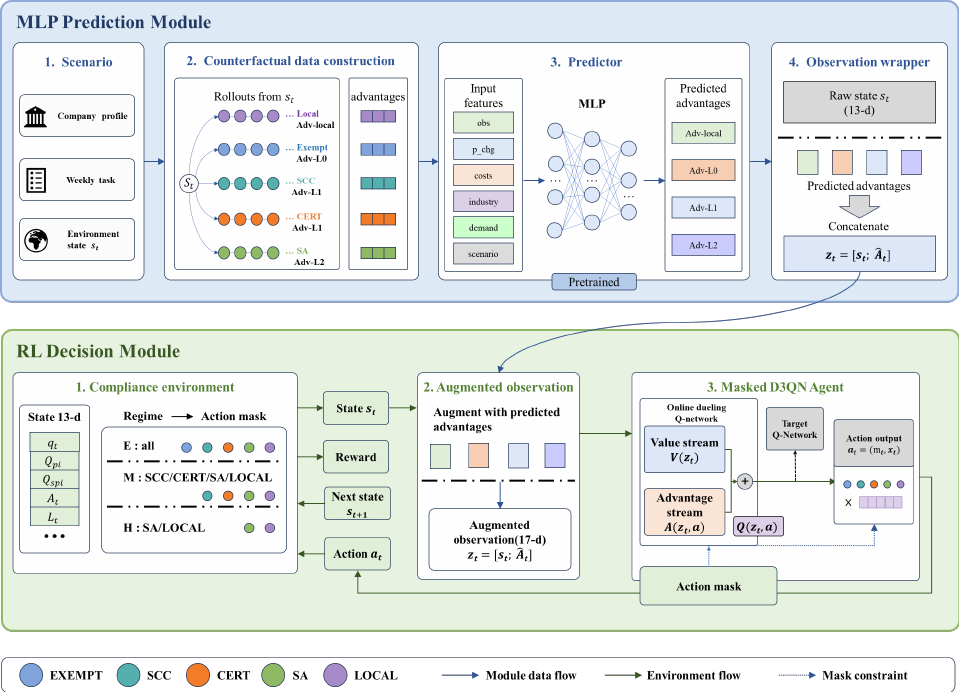}
\caption{Two-stage design of the solution method}
\label{f:fig2}\end{figure}

The myopic single-period problem under a given export mechanism admits an explicit solution, whereas the full sequential problem does not because of intertemporal state coupling. A current decision simultaneously changes the cumulative export volumes, regulatory friction, and credential state, thereby altering the compliance boundary, cost structure, and feasible action set in subsequent weeks. The state also contains continuous variables and evolves endogenously with the firm's decision history, making exact dynamic programming computationally impractical. This paper therefore adopts a two-stage masked reinforcement learning design. First, the long-run advantages of the four strategic paths are estimated offline. A masked learner then incorporates these predicted path-advantage signals into its observation and decides week by week within the legal action set. These signals provide forward-looking information for policy learning and subsequently serve as readable decision support signals, helping managers understand why the policy favors local processing, exempt export, Level-1 credential export, or Level-2 credential export in different states. Fig.~\ref{f:fig2} presents the overall design.

\subsection{Counterfactual path-prediction module}

The counterfactual path-prediction module outputs numerical advantage signals for four strategic paths. Each summarizes the long-run reward increment associated with taking a representative action for a given path in the current week and then reverting to a default continuation policy for the remaining horizon. The four paths are local processing, LOCAL; exempt export, EXPORT-L0 (EXEMPT); Level-1 credential export, EXPORT-L1 (SCC/CERT); and Level-2 credential export, EXPORT-L2 (SA), abbreviated LOCAL, L0, L1, and L2 below. Because the long-run consequence of the current path choice depends on subsequent actions, a deterministic continuation policy is first specified as the reference.

The reference is the default continuation policy $\pi_0$. In tier $E$, it selects full-volume EXEMPT; in tier $M$, it selects full-volume SCC. In tier $H$, it selects full-volume SA if the firm already holds the Level-2 credential and local processing otherwise. Because $\pi_0$ is fixed and deterministic, the corresponding future rollouts and training labels can be generated offline before policy training. Using $\pi_0$ as the reference, the rollout reward of any candidate first action $a$ is defined as the discounted reward from taking $a$ in the current week and following $\pi_0$ thereafter:
\begin{equation}
G(s_t,a)=R(s_t,a)+\sum_{j=1}^{T-1-t}\gamma^{\,j}\,R\!\big(s_{t+j},\,\pi_0(s_{t+j})\big),
\label{e:eq22}\end{equation}
where $R(\cdot)$ is the single-period reward, $\gamma=0.99$ is the label discount factor, and $T=52$ is the length of the compliance year.

The counterfactual advantage of path $k$ is defined as the long-run reward increment from locking into path $k$ this week relative to following the default policy this week:
\begin{equation}
\Lambda^k_t=\Big(G(s_t,a^k)-G\big(s_t,\pi_0(s_t)\big)\Big)\,\mathbb{1}\!\big[a^k\in\mathcal{A}_{\mathrm{legal}}(s_t)\big],\qquad k\in\{\text{LOCAL},\text{L0},\text{L1},\text{L2}\},
\label{e:eq23}\end{equation}
where the representative action $a^k$ assigns zero transfer volume to local processing and full transfer volume to each of the three export paths. The indicator in Eq.~\eqref{e:eq23} removes the paths that are illegal in the current week. The four path advantages are stacked into the vector $\Lambda_t=(\Lambda^{\text{LOCAL}}_t,\Lambda^{\text{L0}}_t,\Lambda^{\text{L1}}_t,\Lambda^{\text{L2}}_t)$, which forms the four-dimensional signal subsequently supplied to the online policy. By construction, the advantage of the path containing the default action $\pi_0(s_t)$ is zero, so the exempt-export advantage is zero wherever that path is legal, and Eq.~\eqref{e:eq23} sets illegal paths to zero as well. The informative content of the signal therefore lies in how the remaining paths deviate from the default policy; the four-dimensional layout is retained so that the observation keeps a fixed length and path order across tiers.

Directly recomputing Eqs.~\eqref{e:eq22}--\eqref{e:eq23} at every decision step during policy training would require repeated rollouts over the remaining horizon under $\pi_0$. The four path advantages are therefore approximated using a predictor trained offline. The input feature $\phi_t$ extends the base state with the environment parameters that affect long-run rewards and with one-hot encodings of the firm's industry, demand band, and scenario type, yielding 32 input dimensions. The predictor uses one-hot encodings because it is trained as an offline regression on a fixed sample, whereas the policy networks consume the compact base observation shared by all learners; the two modules interact only through the four-dimensional signal. The predictor is a three-hidden-layer multilayer perceptron $f_\psi:\mathbb{R}^{32}\to\mathbb{R}^4$ trained on all firm-week samples using masked mean squared error against the path-advantage label $\Lambda_t$:
\begin{equation}
\min_{\psi}\ \frac{1}{N}\sum_{t}\sum_{k}\mathbb{1}\!\big[a^k\in\mathcal{A}_{\mathrm{legal}}(s_t)\big]\,\big(f_\psi(\phi_t)_k-\Lambda^k_t\big)^2.
\label{e:eq24}\end{equation}

The predictions are truncated to $[-c,c]$ ($c=5$) and denoted by $\hat\Lambda_t$. All labels are generated and cached offline before reinforcement learning. Once the predictor is trained, the online stage obtains the signals through a single forward pass of $f_\psi$, without performing additional path rollouts at decision time.

\subsection{Masked RL module}

The online policy incorporates the predicted path-advantage signals $\hat\Lambda_t$ into its week-by-week decisions. The CPAA augmentation is combined with a masked D3QN (Dueling Double DQN) learner, yielding D3QN+CPAA, which serves as the subject of the subsequent decision support analysis. D3QN is selected as the base algorithm because it combines the Double and Dueling improvements of the DQN architecture. Double-DQN-style value learners have also been applied in related decision support research \citep{Zhou25}.

All learners share the same observation, action space, and legal-action-mask setting. For the CPAA-augmented configurations, the policy observation is $z_t=(s_t,\hat\Lambda_t)\in\mathbb{R}^{17}$, which concatenates the 4-dimensional predicted path-advantage vector $\hat\Lambda_t$ with the 13-dimensional base state $s_t$. To fit discrete-action learners, the transfer volume is discretized into an equally spaced grid on $[0,q_t]$; the number of levels is given in Appendix~\ref{a:appendixB}. The policy then selects a composite action $a=(m_t,x_t)$ from the resulting finite action space. The legal-action mask provides the hard feasibility constraint. At every step, it confines the policy to the state-dependent legal action set $\mathcal{A}_{\mathrm{legal}}(s_t)$ and excludes all remaining actions \citep{Huang22,Stolz24}. Export mechanisms below the statutory minimum stringency are therefore unavailable throughout both training and execution, and the reward function contains no violation penalty.

On this basis, D3QN learns a masked action-value function $Q_\theta(z_t,a)$ on the augmented observation. The dueling structure decomposes the action-value estimate into a state-value component and an action-advantage component. The double estimator uses the online network for action selection and the target network for action evaluation, thereby mitigating Q-value overestimation. Both improvements operate within the legal action set. Accordingly, greedy action selection and the maximization in the bootstrap target are restricted to legal actions:
\begin{equation}
a_t=\arg\max_{a\in\mathcal{A}_{\mathrm{legal}}(s_t)} Q_\theta(z_t,a),
\label{e:eq25}\end{equation}
\begin{equation}
y_t=r_t+\gamma\,Q_{\theta^-}\!\Big(z_{t+1},\ \arg\max_{a'\in\mathcal{A}_{\mathrm{legal}}(s_{t+1})} Q_\theta(z_{t+1},a')\Big),
\label{e:eq26}\end{equation}
where $\theta^-$ denotes the target-network parameters and $\gamma$ the discount factor. The network parameters are updated by minimizing the squared temporal-difference error $\big(Q_\theta(z_t,a_t)-y_t\big)^2$.

All baselines follow the same legal-action-mask principle. Among the value-based methods, DQN, Double DQN, and Dueling DQN serve as ablation configurations of D3QN to examine how double estimation and the dueling structure affect policy performance. For the policy-based PPO and A2C actor--critic methods, the logits of illegal actions are set to $-\infty$, after which the action distribution is renormalized over the legal action set. The resulting policy is $\pi_\theta(a\mid z_t)\propto\exp\!\big(h_\theta(z_t)_a\big)\,\mathbb{1}[a\in\mathcal{A}_{\mathrm{legal}}(s_t)]$, and therefore assigns positive probability only to legal actions. The CPAA augmentation is learner-agnostic and can be combined with each of these methods. For every base learner, removing $\hat\Lambda_t$ and restoring the observation to $s_t$ yields the corresponding unaugmented configuration. The paper thus trains unaugmented and +CPAA versions of six base learners to assess the robustness of the main decision support findings and to isolate the effect of the CPAA augmentation. The subsequent interpretability analysis uses D3QN+CPAA as the focal model.

\section{Numerical experiments}\label{s:experiments}

\subsection{Experimental design and evaluation}

The experiments are designed around the three research questions and share the same institutionally anchored environment and evaluation protocol. The first group evaluates the performance of firm-side sequential decisions under a fixed baseline institutional scenario. Masked RL policies are compared with non-learning rule-based baselines to assess whether learned policies improve decision payoffs. The learned policy is further examined as a decision support tool by analyzing how it selects export paths in different states, when it turns to local processing, and when it invests in compliance credentials. The robustness of these behavioral patterns to firm-side parameters is also assessed. The second group turns to the institutional environment and examines how decision boundaries change as firms encounter the assessment-tier constraint more frequently and how the firm's welfare and behavior respond as regulatory friction intensifies.

Both groups use the institutionally anchored environment developed above. Each firm is characterized by a set of intrinsic attributes and a 52-week stream of export tasks. The data type, business type, destination, and demand volume of each task are drawn weekly from calibrated distributions. The compliance structure is determined by the cumulative-volume thresholds and export-routing logic prescribed by the Provisions on Promoting and Regulating the Cross-Border Flow of Data \citep{CAC24}, which determine the legal mechanisms and minimum compliance stringency for each state week by week. The legal constraints are therefore traceable to statutory rules, while the economic parameters are calibrated from empirical evidence rather than selected arbitrarily. Because real regulatory regimes cannot readily be varied under controlled conditions in observational data, the institution-level counterfactual comparisons rely on simulation. Evaluating sequential policies on simulated scenario libraries is also common in operations management and RL-based decision support research \citep{Zhou25,Gijsbrechts22,Harsha25}. Accordingly, the scenario library comprises 3000 training firms, 300 validation firms, and 300 test firms. The validation set is used for model selection, and the test set is reserved for final evaluation. All reported results are averaged over five random seeds.

\subsection{Policy performance}

\begin{figure}[H]
\includegraphics[width=0.7\textwidth,page=3]{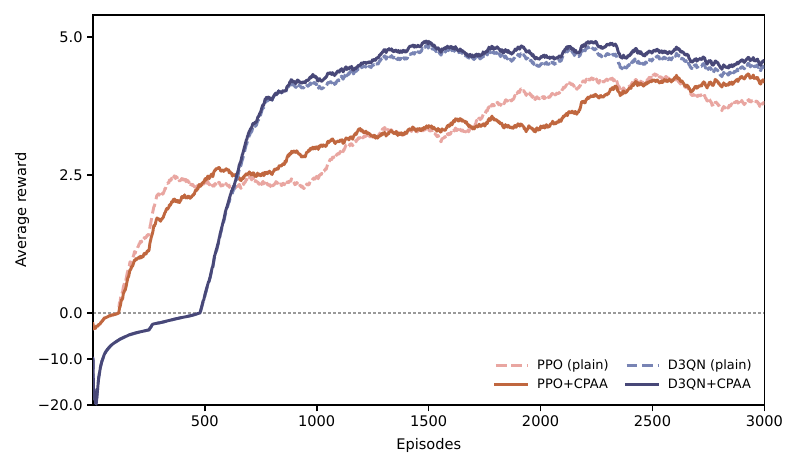}
\caption{Average training-reward trajectories of the four masked RL configurations}
\label{f:fig3}\end{figure}

This section evaluates masked reinforcement learning on the sequential decision problem by comparing the learned policies with non-learning baselines. The Minimum-compliance policy always transfers through the minimum compliance stringency, representing passive adherence to the minimum statutory requirement. The Always-LOCAL policy always chooses local processing, representing the conservative strategy of abandoning cross-border transfer to avoid transfer-related institutional constraint. The learning methods comprise six masked reinforcement learners, including policy-based actor--critic methods and the value-based DQN family. Each learner is trained both without and with counterfactual path-advantage augmentation to examine how the base learner and the CPAA augmentation affect policy performance and behavioral patterns. The hyperparameters, including learning rate, replay capacity, batch size, and discount factor, were tuned through repeated experiments; the final settings are reported in Appendix~\ref{a:appendixB} (Table~\ref{t:tableB2}). Augmented and unaugmented configurations use the same hyperparameters except for the observation dimension. Fig.~\ref{f:fig3} shows the average-reward trajectories over 3000 training episodes for four representative configurations: PPO and D3QN, each with and without CPAA augmentation. Overall, the D3QN models converge faster and stabilize at higher reward levels, whereas the PPO models converge more slowly and reach lower final rewards.

Table~\ref{t:table3} reports the rewards and behavioral metrics of all models on the test set. Rewards are reported as the mean $\pm$ standard deviation over five random seeds. The behavioral metrics are the shares of the four response-path classes (EXEMPT, SCC/CERT, SA, LOCAL) across all weekly decisions, while the illegal column reports the share of attempted illegal actions. The two non-learning policies exhibit clear limitations. The Always-LOCAL policy forgoes the business value of cross-border transfer entirely and earns the lowest reward. The Minimum-compliance policy always satisfies the minimum compliance stringency but lacks forward-looking adjustment to cumulative export volumes, regulatory friction, or credential state; its reward remains below that of most learned policies. These results suggest that the firm-side sequential decision problem is not well served by a single rigid rule and instead requires state-dependent coordination among cross-border transfer, local processing, and credential investment.

\begin{table}[H]
\caption{Returns and behavioral metrics of all policies on the held-out test set}
\begin{tabular*}{\textwidth}[]{@{\extracolsep\fill}lcccccc}
\toprule
Method & Reward & EXEMPT & SCC/CERT & SA & LOCAL & Illegal\\
\midrule
Always-LOCAL & $-7.019\pm0.13$ & $0.000$ & $0.000$ & $0.000$ & $1.000$ & $0$\\
Minimum-compliance & $1.297\pm0.09$ & $0.488$ & $0.235$ & $0.277$ & $0.000$ & $0$\\
PPO & $3.862\pm0.79$ & $0.353$ & $0.337$ & $0.147$ & $0.163$ & $0$\\
PPO+CPAA & $4.184\pm0.30$ & $0.363$ & $0.353$ & $0.167$ & $0.117$ & $0$\\
A2C & $1.916\pm0.92$ & $0.282$ & $0.261$ & $0.457$ & $0.000$ & $0$\\
A2C+CPAA & $2.126\pm0.45$ & $0.470$ & $0.193$ & $0.337$ & $0.000$ & $0$\\
DQN & $4.631\pm0.22$ & $0.305$ & $0.315$ & $0.163$ & $0.217$ & $0$\\
DQN+CPAA & $4.389\pm0.16$ & $0.278$ & $0.362$ & $0.148$ & $0.212$ & $0$\\
Double DQN & $4.559\pm0.20$ & $0.283$ & $0.353$ & $0.163$ & $0.201$ & $0$\\
Double DQN+CPAA & $4.632\pm0.06$ & $0.318$ & $0.312$ & $0.148$ & $0.223$ & $0$\\
Dueling DQN & $4.632\pm0.20$ & $0.308$ & $0.305$ & $0.180$ & $0.206$ & $0$\\
Dueling DQN+CPAA & $4.630\pm0.33$ & $0.317$ & $0.316$ & $0.156$ & $0.211$ & $0$\\
D3QN & $4.650\pm0.22$ & $0.302$ & $0.313$ & $0.156$ & $0.229$ & $0$\\
\textbf{D3QN+CPAA} & $\mathbf{4.959\pm0.18}$ & $0.360$ & $0.289$ & $0.136$ & $0.215$ & $0$\\
\bottomrule
\end{tabular*}
\note{Bold marks the top-performing policies: those whose mean reward lies within one standard deviation of the best-performing policy.}
\label{t:table3}\end{table}

Among the learned policies, the value-based DQN family generally achieves higher rewards and lower variation than PPO and A2C, and D3QN+CPAA attains the highest average reward. The effect of CPAA nevertheless varies across base learners. CPAA is therefore not interpreted as a universally performance-enhancing module but as a source of forward-looking counterfactual path-advantage signals that can support both policy learning and interpretation. The learned policies also differ behaviorally from both rule-based baselines, making state-dependent choices among SCC/CERT, SA, and LOCAL rather than mechanically following minimum compliance or full localization. As expected from the hard-constraint design, all masked policies attempt zero illegal actions on the test set. This feasibility is enforced by the legal action sets generated by the minimal compliance mapping and the action mask, rather than by violation penalties in the reward. Overall, the learned policies provide a basis for examining the localization and credential-investment decisions analyzed next.

\subsection{Deeper insights into the system's results}

\begin{figure}[H]
\includegraphics[width=0.7\textwidth,page=4]{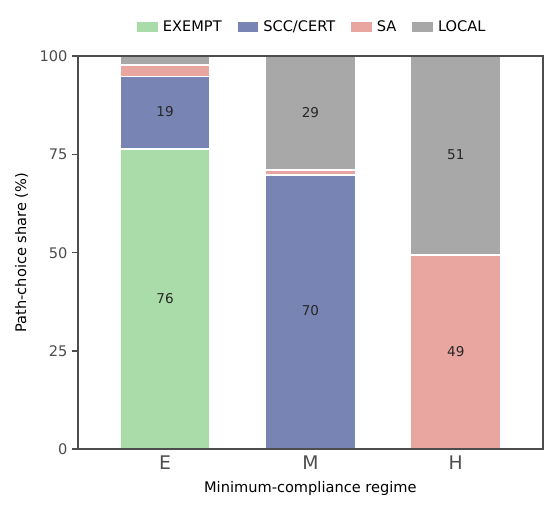}
\caption{Path-choice distributions across minimum compliance stringency tiers}
\label{f:fig4}\end{figure}

Fig.~\ref{f:fig4} first presents the distribution of path choices within each minimum compliance stringency tier. In the exemption tier $E$, the policy relies primarily on exempt transfer; in the standard tier $M$, it moves to the standard contract or protection certification; in the assessment tier $H$, the policy divides its choices approximately evenly between the security assessment and local processing.

\begin{figure}[H]
\includegraphics[width=0.7\textwidth,page=5]{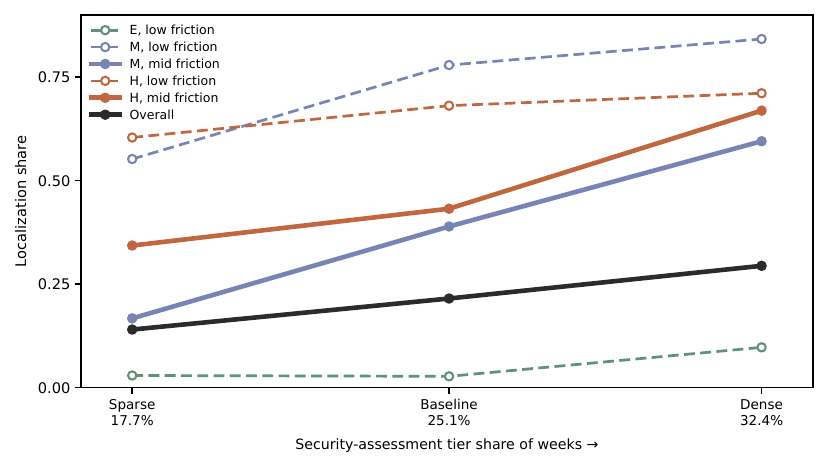}
\caption{Changes in the localization share of five key states as the share of assessment-tier weeks rises}
\label{f:fig5}\end{figure}

Three scenario libraries are used to examine how changes in the institutional environment shift the firm's decision boundaries. The libraries hold the statutory thresholds, the reward function, and all other environment parameters constant and vary only the composition of the firm population. This variation progressively increases the share of weeks in which export tasks are subject to the assessment-tier ($H$) constraint. The share is measured at library construction by applying the minimal compliance mapping along each firm's generated task sequence under a full-volume export bound, and is therefore a property of the scenario libraries rather than of any learned policy. The corresponding shares are 17.7\%, 25.1\%, and 32.4\%, respectively, hereafter referred to as the sparse, baseline, and dense scenarios. Policies are retrained separately on each library under the identical configuration and evaluated on that library's held-out test firms. More frequent exposure to the assessment tier means that firm-side export activity is more often subject to the highest compliance stringency. The manipulation therefore represents tighter operational exposure to the existing regulatory regime while leaving the statutory rules themselves unchanged. Against this background, the analysis examines how the localization boundary shifts as the share of assessment-tier weeks increases. Table~\ref{t:table4} characterizes localization behavior by the state occupied by the firm. Each state is defined along three dimensions: the current minimum compliance stringency, the accumulated regulatory friction stock, and the current export demand. The stringency tiers are exemption $E$, standard $M$, and assessment $H$. The regulatory friction $A_t$ is classified as low ($A_t<0.15$), medium ($0.15\le A_t<0.4$), and high ($A_t\ge 0.4$). The export demand is measured by the ratio of weekly demand to the reference scale, $q_t/q_{\mathrm{ref}}$, and classified as low (ratio $<0.3$), medium ($0.3\le$ ratio $<0.8$), and high (ratio $\ge 0.8$). Table~\ref{t:table4} reports the localization share of each state across the three scenarios, while Fig.~\ref{f:fig5} traces five key states as the prevalence of assessment-tier weeks increases.

\begin{table}[H]
\caption{Localization share by state}
\begin{tabular*}{\textwidth}[]{@{\extracolsep\fill}lccc}
\toprule
State (tier $\cdot$ friction $\cdot$ demand) & Sparse & Baseline & Dense\\
\midrule
$E$ $\cdot$ low $\cdot$ low & $0.029$ & $0.027$ & $0.097$\\
$M$ $\cdot$ low $\cdot$ low & $0.552$ & $0.779$ & $0.842$\\
$H$ $\cdot$ low $\cdot$ low & $0.604$ & $0.681$ & $0.711$\\
$M$ $\cdot$ medium $\cdot$ low & $0.167$ & $0.389$ & $0.595$\\
$H$ $\cdot$ medium $\cdot$ low & $0.343$ & $0.432$ & $0.669$\\
All other states & $\approx 0$ & $\approx 0$ & $\approx 0$\\
\bottomrule
\end{tabular*}
\label{t:table4}\end{table}

Local processing concentrates in states where the business value of a small lawful transfer does not cover its compliance cost. Assessment-tier compliance costs are largely driven by credential acquisition and maintenance, which are difficult to amortize over small transfer volumes. The policy therefore turns to local processing frequently in non-exempt, low-demand states, while localization remains rare when demand is high or an exemption applies. From the sparse to the dense scenario, the overall localization share rises from $0.140$ to $0.294$. The largest increases occur in critical medium friction states. In the standard tier under medium friction, for example, the share rises from $0.167$ to $0.595$, extending local processing into states previously dominated by export. By contrast, localization in the exemption tier changes much less. This pattern is consistent with the empirical evidence from the GDPR that marginal data activities are reduced first when compliance costs increase \citep{Jia21,Johnson23}.

\subsection{Interpretable decision signals and decision trees}

The learned policy converts into interpretable managerial guidance through two complementary channels: directly readable counterfactual path signals and shallow decision trees distilled ex post from policy behavior.

The counterfactual path advantages are taken directly from the policy's input. The four components correspond to local processing and the three export paths. Each measures the long-run reward increment from taking the representative action associated with a strategic path in the current week, relative to the default continuation policy $\pi_0$. Signals for paths that are illegal in the current week are set to zero, while the legal-action mask separately excludes the corresponding actions from the policy's choice set. The signal $\hat\Lambda_t$ therefore quantifies the value of the remaining legal options; it enters the augmented observation $z_t=(s_t,\hat\Lambda_t)$ for policy computation and can be used directly by the compliance manager.

\begin{figure}[H]
\includegraphics[width=0.7\textwidth,page=6]{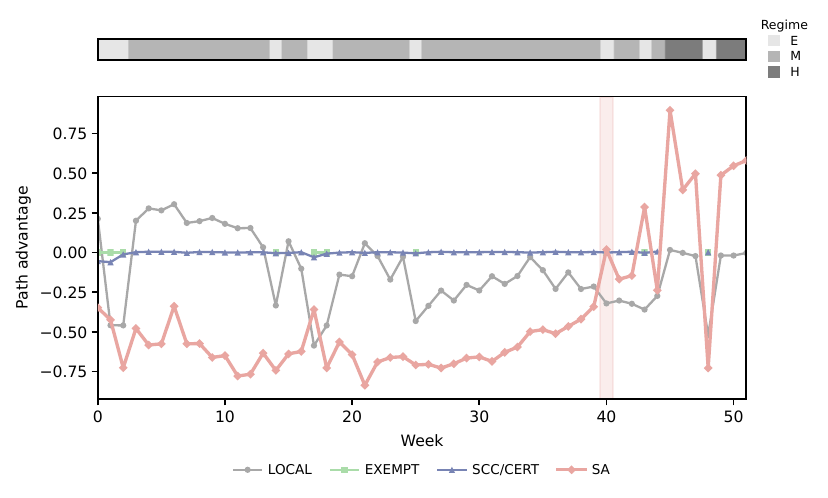}
\caption{Week-by-week evolution of the four counterfactual path advantages of a representative firm over a compliance year}
\label{f:fig6}\end{figure}

Fig.~\ref{f:fig6} traces the signals of a representative firm over one compliance year. The strip at the top indicates the minimum compliance stringency in each week, and only signals for currently legal paths are plotted. The exemption advantage appears only in weeks when the firm is in tier $E$ and is identically zero, while the standard-contract advantage is unavailable in assessment-tier weeks. The security-assessment advantage is markedly negative throughout the standard-tier weeks, indicating that choosing the security assessment in those states constitutes over-compliance whose benefit cannot cover its cost. As cumulative export volume approaches the assessment-tier threshold, the security-assessment advantage becomes positive around week 40, preceding the policy's actual switch to the security assessment in week 45.

On the localization path, the localization share is about $60\%$ in weeks when the localization advantage is positive and about $10\%$ in the remaining weeks, against an overall level of about $21.5\%$ that matches the localization share reported in Table~\ref{t:table3}. A positive localization advantage therefore aligns with actual switches to local processing. A similar relationship appears in credential investment: among firms that actually switch to the security assessment, the security-assessment advantage turns non-negative in or before the adoption week in about $71\%$ of the cases. Across weeks in which several legal paths remain available, the legal path with the highest advantage matches the policy's actual choice approximately $73\%$ of the time, with the match rate lowest in the assessment tier.

Although the masked RL policy provides executable week-by-week recommendations, its underlying neural network is difficult for a compliance manager to execute manually or inspect for generalizable decision rules. To express the policy in a form that can be communicated to managers and auditors, the trained policy is therefore approximated using shallow decision trees, following established policy-distillation practice \citep{BastaniO18,Glanois24,Zhou25}. The D3QN+CPAA policy is first executed on the test scenarios, and its weekly states and selected actions are recorded. Two decision trees are then fitted: one characterizes when the policy turns to local processing, and the other characterizes whether and when it invests in a credential. Both trees have a maximum depth of 3, so a recommendation can be reached in at most three decision checks. The candidate split features are observable state variables and exclude the counterfactual path advantages; the fitted trees split on data sensitivity type, compliance stringency tier, credential holding, weekly demand, friction stock, and cumulative export volume. The trees therefore characterize the policy at the input--output level and complement the real-time signals above. The thresholds on continuous features are learned directly from the policy rather than generic rules of thumb.

\begin{figure}[H]
\subcaptionbox{Localization tree\label{f:tree_local}}{\includegraphics[width=0.9\textwidth,page=7]{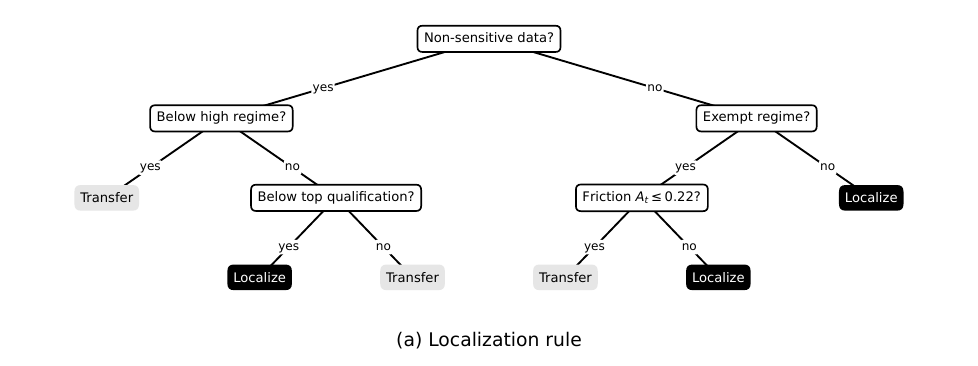}}\\
\subcaptionbox{Investment tree\label{f:tree_invest}}{\includegraphics[width=0.9\textwidth,page=8]{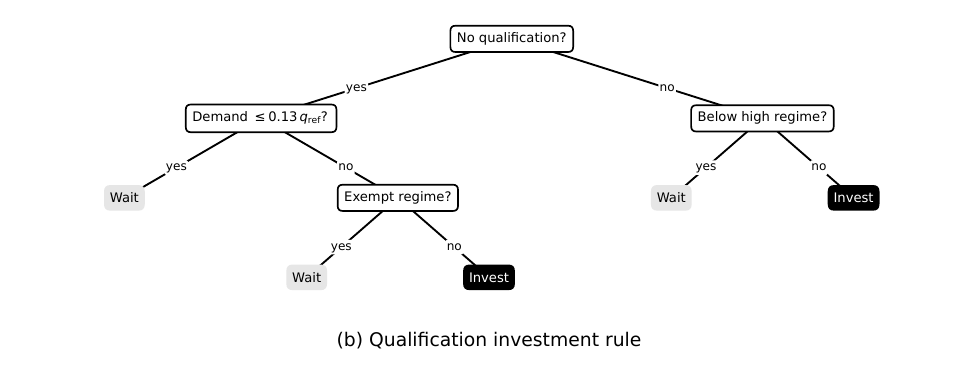}}
\caption{Shallow decision trees distilled from the D3QN+CPAA policy}
\label{f:fig7}\end{figure}

The localization tree in Fig.~\ref{f:tree_local} uses data sensitivity as its root node. For non-sensitive data, it recommends export throughout the exemption and standard tiers and turns to local processing only when the state enters the assessment tier and the firm does not yet hold the highest-level credential. For sensitive or important data, it turns to local processing from the standard tier upward. Even in the exemption tier, it recommends export only when the friction stock is below $0.22$ and switches to local processing above that threshold. The investment tree in Fig.~\ref{f:tree_invest} uses credential holding as its root node. For firms without a credential, it recommends immediate acquisition once weekly demand exceeds $13\%$ of the reference scale and the state has entered the standard tier or above. Firms that already hold a credential upgrade upon entering the assessment tier and otherwise wait. The two trees thus translate the questions of when to localize, whether and when to invest into threshold-based, actionable rules. They also show that the policy tends to front-load credential acquisition within the compliance year.

Following these rules, a compliance manager can reproduce the learned policy's weekly decisions without running the network, with fidelities of $92.3\%$ for the localization tree and $97.6\%$ for the investment tree. The deviations concentrate in a few critical states where the learned policy conditions on friction stock and cumulative export volume more finely than a depth-3 tree can represent. This difference reflects the trade-off between interpretability and fidelity. Taken together, the real-time path-advantage signals, executable shallow trees, and underlying policy provide a complementary framework of decision support: forward-looking value comparisons, auditable decision rules, and finer-grained recommendations in critical states.

\subsection{Robustness to firm-side parameters}

The preceding decision support conclusions are obtained under one calibration of the environment, and their dependence on particular parameter values requires examination. Beyond the institutional environment, the conclusions may also be affected by the firm's own economic fundamentals. This section varies the curvature of the business-value function and the Level-2 credential-acquisition cost separately to assess whether the main behavioral patterns persist.

The value-curvature parameter $\beta$ governs the shape and marginal gains of the business-value function, and the Level-2 acquisition cost $F(2)$ captures the investment cost of obtaining the corresponding credential. Because changes in $\beta$ alter the level of realized rewards, raw rewards are not directly comparable across parameter settings. The analysis therefore focuses on behavioral metrics directly related to the main decision support conclusions.

\begin{figure}[H]
\includegraphics[width=0.8\textwidth,page=9]{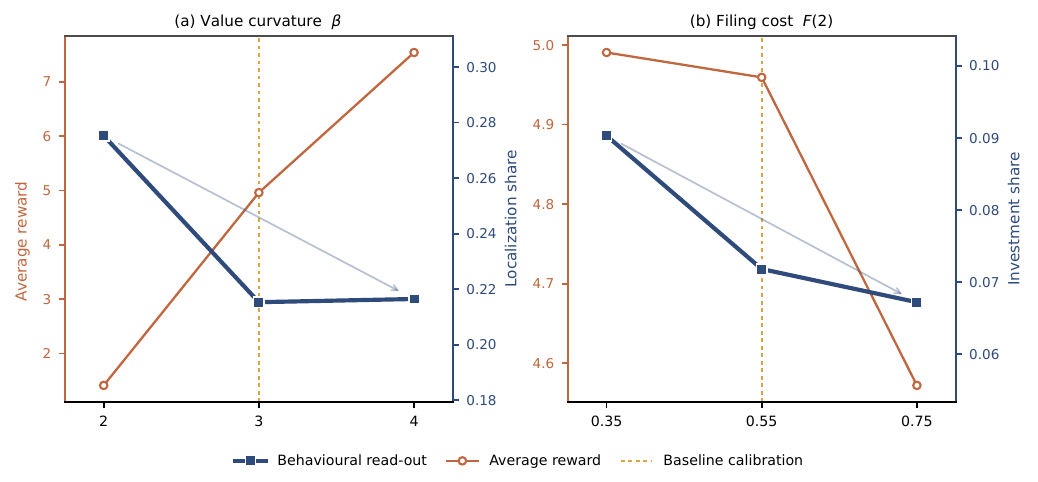}
\caption{Behavioral responses of the policy to firm-side parameter perturbations. (a) Localization share as the value curvature $\beta$ varies; (b) new-credential share as the Level-2 acquisition cost $F(2)$ varies}
\label{f:fig8}\end{figure}

Table~\ref{t:table5} and Fig.~\ref{f:fig8} summarize the behavioral responses to the firm-side parameter perturbations. The qualitative patterns remain stable, while the magnitude of the responses changes. Fig.~\ref{f:fig8}a shows that, as $\beta$ increases, the localization share declines from its highest level in the low-$\beta$ setting to approximately the baseline and then remains broadly stable. Higher transfer value weakens the firm's tendency to turn to local processing but does not eliminate it within the tested range. Therefore, local processing is not an artifact of undervaluing cross-border transfer but a structural response that persists under the given compliance costs and institutional constraints. Fig.~\ref{f:fig8}b shows that as $F(2)$ increases, the share of new credential acquisitions declines but remains positive. A higher acquisition cost therefore reduces the frequency of investment without changing the firm's basic pattern of arranging credentials early in the compliance year. Overall, the firm-side parameter perturbations mainly affect the magnitude of localization and investment responses, while the behaviors underlying the main decision support findings persist across the tested settings.

\begin{table}[H]
\caption{Robustness to firm-side economic calibration}
\begin{tabular*}{\textwidth}[]{@{\extracolsep\fill}lp{3.6cm}ccc}
\toprule
Parameter (low / baseline / high) & Behavioral readout & Low & Baseline & High\\
\midrule
Value curvature $\beta$ ($2.0/3.0/4.0$) & Localization share & $0.275\pm0.022$ & $0.215\pm0.012$ & $0.216\pm0.035$\\
Acquisition cost $F(2)$ ($0.35/0.55/0.75$) & New-credential share & $0.090\pm0.015$ & $0.072\pm0.005$ & $0.067\pm0.005$\\
\bottomrule
\end{tabular*}
\label{t:table5}\end{table}

\subsection{Welfare and behavioral effects of regulatory friction}

The institutionally anchored decision environment supports not only the analysis of firm-level policies but also the examination of how changes in the regulatory policy affect firms' payoffs and behavior. This section treats the persistent-friction weight $\kappa_A$, which determines the cost borne by the firm as regulatory friction accumulates, as a continuous representation of regulatory strictness. It then examines how the firm's expected reward, localization share, and cross-border transfer volume change as this friction intensifies. The experiment increases the weight stepwise from its calibrated baseline of $0.5$ to $2$, holding all other parameters constant.

The effect of $\kappa_A$ can first be explained from the model structure. When comparing the single-period net payoffs of cross-border transfer and local processing within a week, the current friction cost $\kappa_A A_t$ depends only on the beginning-of-period friction stock $A_t$ and is therefore common to both current choices. Let $x^{\star}$ denote the transfer volume that maximizes the single-period reward under a fixed export mechanism. Its value depends on the business-value parameters and the marginal transfer costs and likewise does not involve $\kappa_A$. The single-period payoff difference between cross-border transfer and local processing can therefore be written as:
\begin{equation}
\Delta R = b(\tau)(1+\mu)\,g\!\bigl(x^{\star}/q_{\mathrm{ref}}\bigr) - \bigl[F\,\mathbf{1}\{\ell(m)>L_t\}+\mathrm{maint}\bigr] - \bigl(c_{\mathrm{var}}+\kappa_\sigma\bigr)\,x^{\star}/q_{\mathrm{ref}}.
\label{e:eq27}\end{equation}

The formula does not contain $\kappa_A$, so the persistent-friction weight does not directly change the payoff difference between cross-border transfer and local processing within the week. Its effect is instead sequential: the current transfer decision changes the future friction stock, and a larger $\kappa_A$ increases the cost associated with that stock in subsequent periods. The parameter therefore operates primarily through the future consequences of current transfer decisions rather than through the static payoff comparison in the current week.

\begin{figure}[H]
\includegraphics[width=0.7\textwidth,page=10]{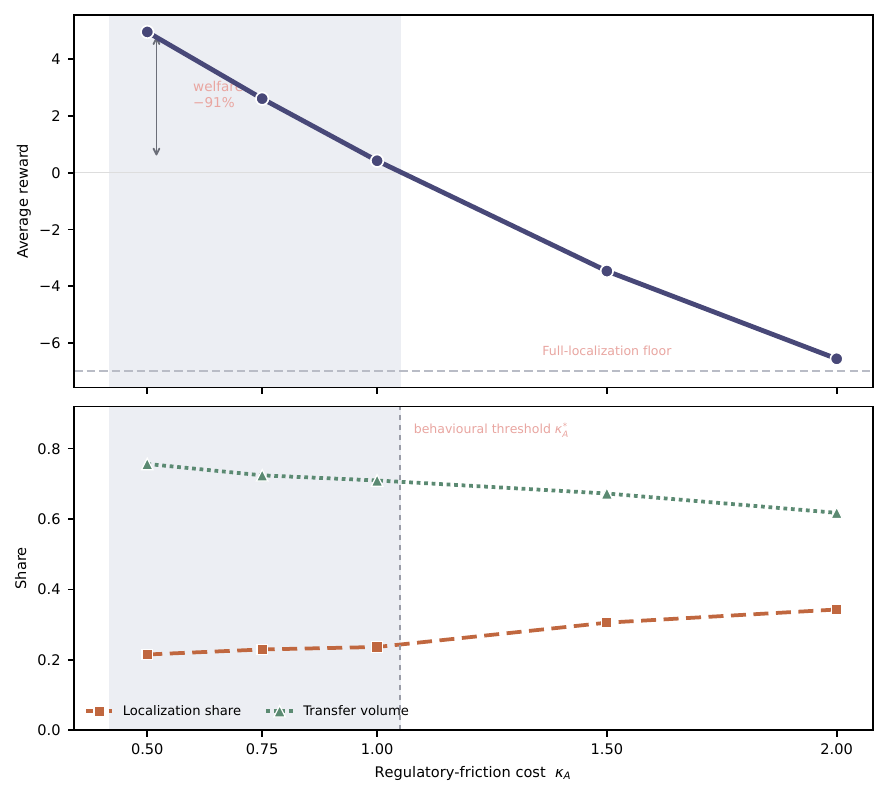}
\caption{Welfare (top) and behavioral responses (bottom) of the policy under different persistent-friction weights $\kappa_A$}
\label{f:fig9}\end{figure}

Fig.~\ref{f:fig9} shows how expected reward and behavioral responses change under different $\kappa_A$ values. Over the initial range, expected reward already declines markedly while the localization share and transfer volume change little. The firm thus absorbs higher persistent-friction costs without immediately changing its observable behavior. Only after the friction intensity crosses a behavioral threshold does the previous decision balance change visibly: the policy increases local processing and reduces cross-border transfer volume. Under the simulation settings examined here, stronger regulatory friction therefore produces a pattern in which expected reward declines before observable behavior changes. This pattern is consistent with the core mechanism of lumpy adjustment theory: when adjustment involves sunk or sequential costs, optimal behavior can remain unchanged over a range of conditions and then adjust more sharply once a threshold is crossed \citep{Baley26}.

This finding has implications for the assessment of regulatory costs. If the effects of data-governance measures are evaluated only through observable behavioral indicators, such as cross-border transfer volume, localization shares, or firm relocation, the resulting assessment may understate costs already borne by firms. Expected rewards can decline before a visible behavioral response emerges. This risk is particularly relevant at intermediate levels of regulatory friction, where firms exhibit little localization or transfer compression even though persistent costs have already reduced their expected payoffs.

Assessments of the economic effects of cross-border data governance should therefore consider not only whether firms visibly change their behavior but also how regulatory friction affects their net payoffs, unmet transfer demand, and credential-investment pressure over time. The results may also help explain why the economic costs of data-governance measures are difficult to infer directly from behavioral indicators alone \citep{OECDWTO25}.

\section{Conclusions and future directions}\label{s:ccl}

From the perspective of a data-exporting firm, this study examines the sequential compliance decision problem under the data-export regime in which the firm's own history continuously reshapes its future decision environment. The firm must decide each week how much data to transfer, whether to transfer data across borders or turn to local processing, and whether and when to invest in a compliance credential. Because current decisions change cumulative export volumes, regulatory friction, and credential holdings, they also affect future compliance requirements, feasible actions, and costs. Firm-side cross-border data compliance is therefore better understood not as a one-off legal judgment but as a dynamic decision problem under institutionally determined and history-dependent constraints.

To address this problem, the paper develops the institutionally anchored decision support system. The minimal compliance mapping converts the data-export routing logic of the Provisions on Promoting and Regulating the Cross-Border Flow of Data \citep{CAC24} into computable minimum compliance stringency requirements and state-dependent legal action sets. Compliance requirements therefore define the feasible region of the decision problem as hard constraints rather than entering the reward function as penalties. On this basis, counterfactual path-advantage augmentation (CPAA) estimates the long-run reward increments associated with local processing and alternative export paths relative to a default continuation policy and incorporates the predicted signals into the policy observation. Masked reinforcement learning then learns policies within the legal action set. Among the compared learners and two non-learning rules, D3QN+CPAA attains the highest average reward, although the performance effect of CPAA varies across base learners.

The system provides decision support through two complementary forms of interpretability. First, the predicted counterfactual path advantages expose forward-looking differences in the estimated long-run value of alternative strategic paths, helping managers understand why the policy favors local processing, standard contract or protection certification, or security assessment in particular states. Second, shallow decision trees distill the learned policy into executable rules based on observable state variables. The depth-3 localization and credential-investment trees reproduce the corresponding policy decisions with fidelities of $92.3\%$ and $97.6\%$, respectively, allowing managers to approximate the learned policy without executing the neural network.

The resulting policy answers the first two research questions. It does not mechanically follow the minimum-compliance path or rely uniformly on local processing. Instead, it adjusts its decisions to the firm's current compliance stringency, export demand, cumulative history, regulatory friction, and credential state. Local processing concentrates in states where the business value of small lawful transfers does not cover their compliance costs, and the localization boundary shifts as firms encounter high-stringency constraints more frequently. Credential acquisition, meanwhile, tends to be front-loaded within the compliance year. Together, these findings show how the learned policy converts a complex sequential compliance problem into forward-looking signals and threshold-based managerial rules.

The institutionally anchored environment also supports examining how changes in the regulator's own policy parameters affect firm behavior and net payoffs. Treating the persistent-friction weight as a model-based continuous representation of regulatory strictness reveals an absorb-then-adjust pattern under the simulation settings examined here. As regulatory friction intensifies, expected reward initially declines while observable indicators such as the localization share and cross-border transfer volume change little. Only after a behavioral threshold is crossed does the policy visibly increase local processing and reduce transfer volume. This pattern is consistent with the mechanism of lumpy adjustment theory, whereby behavior may remain unchanged over a range of conditions when adjustment involves sunk or sequential costs and then change more sharply once a threshold is reached \citep{Baley26}. This result has implications for the assessment of regulatory costs. When declines in firm payoffs precede observable behavioral responses, assessments based only on transfer volumes, localization shares, or other visible adjustments may understate costs that firms are already bearing. The risk is particularly relevant at intermediate levels of regulatory friction, where expected rewards have declined but behavior has not yet adjusted substantially. Evaluations of cross-border data governance should therefore consider not only whether firms change their observable behavior but also how regulatory friction affects their net payoffs, unmet transfer demand, and credential-investment pressures over time. More generally, the findings suggest that the effects of institutional constraints may emerge in firm payoffs before they become visible in aggregate behavior.

The system is not intrinsically limited to China's data-export regime, although transferring it to another setting requires the institutional mapping and economic environment to be reconstructed for that setting. For another jurisdiction, the minimal compliance mapping can be rewritten to represent the relevant condition-triggered compliance mechanisms and their priority structure. More broadly, the system may apply to rule-based compliance problems in which feasible actions are determined by formal requirements and evolve with the decision-maker's own history. Potential settings include total-quantity controls on emission permits and tiered reporting requirements for financial institutions. In such settings, the general sequence of rule mapping, constrained policy learning, and policy distillation may provide a reusable decision support architecture.

This study has two main limitations, each suggesting a direction for future research. First, the simulation environment is calibrated to the data-export regulation and existing economic evidence rather than estimated from proprietary firm-level operational data. Practical deployment would require data-collection and calibration mechanisms that record historical export activity, compliance costs, and credential states and estimate the corresponding environment parameters. Future work could also incorporate online learning for demand distributions that are initially unknown or change over time, while continuing to enforce the applicable compliance boundary as a hard constraint. Second, the analysis adopts the perspective of a single firm. It does not model strategic interactions among firms subject to the same regime, nor does it endogenize the regulator's choice of policy strictness. Future research could extend the system to multi-agent environments that jointly represent inter-firm strategic behavior and regulatory decisions. Such extensions would make it possible to examine whether the separation between declines in firm payoffs and observable behavioral responses persists once market interactions and policy responses are determined endogenously.

\bibliography{paper.bib}

\begin{thebibliography}{36}
\newcommand{\enquote}[1]{``#1''}
\providecommand{\natexlab}[1]{#1}
\providecommand{\url}[1]{\texttt{#1}}

\bibitem[{Aaronson and Leblond(2018)}]{Aaronson18}
Aaronson, Susan~Ariel and Patrick Leblond. 2018.
\newblock \enquote{Another Digital Divide: The Rise of Data Realms and its
  Implications for the {WTO}.}
\newblock \emph{Journal of International Economic Law} 21~(2): 245--272.
\newblock \url{https://doi.org/10.1093/jiel/jgy019}.

\bibitem[{Abbasi et~al.(2012)Abbasi, Albrecht, Vance, and Hansen}]{Abbasi12}
Abbasi, Ahmed, Conan Albrecht, Anthony Vance, and James Hansen. 2012.
\newblock \enquote{MetaFraud: A Meta-Learning Framework for Detecting Financial
  Fraud.}
\newblock \emph{MIS Quarterly} 36~(4): 1293--1328.
\newblock \url{https://doi.org/10.2307/41703508}.

\bibitem[{Acemoglu et~al.(2022)Acemoglu, Makhdoumi, Malekian, and
  Ozdaglar}]{Acemoglu22}
Acemoglu, Daron, Ali Makhdoumi, Azarakhsh Malekian, and Asu Ozdaglar. 2022.
\newblock \enquote{Too Much Data: Prices and Inefficiencies in Data Markets.}
\newblock \emph{American Economic Journal: Microeconomics} 14~(4): 218--256.
\newblock \url{https://doi.org/10.1257/mic.20200200}.

\bibitem[{Baley and Blanco(2026)}]{Baley26}
Baley, Isaac and Andr\'{e}s Blanco. 2026.
\newblock \enquote{The Macroeconomics of Irreversibility.}
\newblock \emph{Review of Economic Studies} p. rdag001.
\newblock \url{https://doi.org/10.1093/restud/rdag001}.

\bibitem[{Bao et~al.(2020)Bao, Ke, Li, Yu, and Zhang}]{Bao20}
Bao, Yang, Bin Ke, Bin Li, Y.~Julia Yu, and Jie Zhang. 2020.
\newblock \enquote{Detecting Accounting Fraud in Publicly Traded {U.S.} Firms
  Using a Machine Learning Approach.}
\newblock \emph{Journal of Accounting Research} 58~(1): 199--235.
\newblock \url{https://doi.org/10.1111/1475-679X.12292}.

\bibitem[{Bastani, Bastani, and Sinchaisri(2026)}]{BastaniH26}
Bastani, Hamsa, Osbert Bastani, and Wichinpong~Park Sinchaisri. 2026.
\newblock \enquote{Improving Human Sequential Decision Making with
  Reinforcement Learning.}
\newblock \emph{Management Science} 72~(1): 733--755.
\newblock \url{https://doi.org/10.1287/mnsc.2022.02455}.

\bibitem[{Bastani, Pu, and Solar-Lezama(2018)}]{BastaniO18}
Bastani, Osbert, Yewen Pu, and Armando Solar-Lezama. 2018.
\newblock \enquote{Verifiable Reinforcement Learning via Policy Extraction.}
\newblock In \emph{Advances in Neural Information Processing Systems 31
  (NeurIPS 2018)}, pp. 2499--2509. Red Hook, NY: Curran Associates.

\bibitem[{Chen et~al.(2023)Chen, Xu, Yu, and Zhang}]{Chen23}
Chen, Ji, Yifan Xu, Peiwen Yu, and Jun Zhang. 2023.
\newblock \enquote{A Reinforcement Learning Approach for Hotel Revenue
  Management with Evidence from Field Experiments.}
\newblock \emph{Journal of Operations Management} 69~(7): 1176--1201.
\newblock \url{https://doi.org/10.1002/joom.1246}.

\bibitem[{Chisam et~al.(2026)Chisam, Moffett, Germann, and
  Palmatier}]{Chisam26}
Chisam, Natalie, Jordan~W. Moffett, Frank Germann, and Robert~W. Palmatier.
  2026.
\newblock \enquote{Privacy Trade-Offs in International Markets.}
\newblock \emph{Journal of International Business Studies}.
\newblock \url{https://doi.org/10.1057/s41267-025-00837-4}.
\newblock Online first 2025.

\bibitem[{{Cyberspace Administration of China (CAC)}(2022)}]{CAC22}
{Cyberspace Administration of China (CAC)}. 2022.
\newblock \enquote{Measures for the Security Assessment of Outbound Data
  Transfers.}
\newblock , Cyberspace Administration of China, Beijing.
\newblock \url{https://www.cac.gov.cn/2022-07/07/c_1658811536396503.htm}.
\newblock Effective 1 September 2022.

\bibitem[{{Cyberspace Administration of China (CAC)}(2023)}]{CAC23}
{Cyberspace Administration of China (CAC)}. 2023.
\newblock \enquote{Measures on the Standard Contract for the Outbound
  Cross-Border Transfer of Personal Information.}
\newblock , Cyberspace Administration of China, Beijing.
\newblock \url{https://www.cac.gov.cn/2023-02/24/c_1678884830036813.htm}.
\newblock Effective 1 June 2023.

\bibitem[{{Cyberspace Administration of China (CAC)}(2024)}]{CAC24}
{Cyberspace Administration of China (CAC)}. 2024.
\newblock \enquote{Provisions on Promoting and Regulating the Cross-Border Flow
  of Data.}
\newblock , Cyberspace Administration of China, Beijing.
\newblock \url{https://www.cac.gov.cn/2024-03/22/c_1712776611775634.htm}.
\newblock Promulgated and effective 22 March 2024.

\bibitem[{Farboodi and Veldkamp(2023)}]{Farboodi23}
Farboodi, Maryam and Laura Veldkamp. 2023.
\newblock \enquote{Data and Markets.}
\newblock \emph{Annual Review of Economics} 15~(1): 23--40.
\newblock \url{https://doi.org/10.1146/annurev-economics-082322-023244}.

\bibitem[{Gijsbrechts et~al.(2022)Gijsbrechts, Boute, Mieghem, and
  Zhang}]{Gijsbrechts22}
Gijsbrechts, Joren, Robert~N. Boute, Jan A.~Van Mieghem, and Dennis~J. Zhang.
  2022.
\newblock \enquote{Can Deep Reinforcement Learning Improve Inventory
  Management? Performance on Lost Sales, Dual-Sourcing, and Multi-Echelon
  Problems.}
\newblock \emph{Manufacturing \& Service Operations Management} 24~(3):
  1349--1368.
\newblock \url{https://doi.org/10.1287/msom.2021.1064}.

\bibitem[{Glanois et~al.(2024)Glanois, Weng, Zimmer, Li, Yang, Hao, and
  Liu}]{Glanois24}
Glanois, Claire, Paul Weng, Matthieu Zimmer, Dong Li, Tianpei Yang, Jianye Hao,
  and Wulong Liu. 2024.
\newblock \enquote{A Survey on Interpretable Reinforcement Learning.}
\newblock \emph{Machine Learning} 113~(8): 5847--5890.
\newblock \url{https://doi.org/10.1007/s10994-024-06543-w}.

\bibitem[{Goldfarb and Tucker(2019)}]{Goldfarb19}
Goldfarb, Avi and Catherine Tucker. 2019.
\newblock \enquote{Digital Economics.}
\newblock \emph{Journal of Economic Literature} 57~(1): 3--43.
\newblock \url{https://doi.org/10.1257/jel.20171452}.

\bibitem[{Harsha et~al.(2025)Harsha, Jagmohan, Kalagnanam, Quanz, and
  Singhvi}]{Harsha25}
Harsha, Pavithra, Ashish Jagmohan, Jayant Kalagnanam, Brian Quanz, and Divya
  Singhvi. 2025.
\newblock \enquote{Deep Policy Iteration with Integer Programming for Inventory
  Management.}
\newblock \emph{Manufacturing \& Service Operations Management} 27~(2):
  369--388.
\newblock \url{https://doi.org/10.1287/msom.2022.0617}.

\bibitem[{Hashmi et~al.(2018)Hashmi, Governatori, Lam, and Wynn}]{Hashmi18}
Hashmi, Mustafa, Guido Governatori, Ho-Pun Lam, and Moe~Thandar Wynn. 2018.
\newblock \enquote{Are We Done with Business Process Compliance: State of the
  Art and Challenges Ahead.}
\newblock \emph{Knowledge and Information Systems} 57~(1): 79--133.
\newblock \url{https://doi.org/10.1007/s10115-017-1142-1}.

\bibitem[{Huang and Ontañón(2022)}]{Huang22}
Huang, Shengyi and Santiago Ontañón. 2022.
\newblock \enquote{A Closer Look at Invalid Action Masking in Policy Gradient
  Algorithms.}
\newblock In \emph{Proceedings of the 35th International Florida Artificial
  Intelligence Research Society Conference (FLAIRS-35)}.
\newblock \url{https://doi.org/10.32473/flairs.v35i.130584}.

\bibitem[{Jia, Jin, and Wagman(2021)}]{Jia21}
Jia, Jian, Ginger~Zhe Jin, and Liad Wagman. 2021.
\newblock \enquote{The Short-Run Effects of the General Data Protection
  Regulation on Technology Venture Investment.}
\newblock \emph{Marketing Science} 40~(4): 661--684.
\newblock \url{https://doi.org/10.1287/mksc.2020.1271}.

\bibitem[{Johnson, Shriver, and Goldberg(2023)}]{Johnson23}
Johnson, Garrett~A., Scott~K. Shriver, and Samuel~G. Goldberg. 2023.
\newblock \enquote{Privacy and Market Concentration: Intended and Unintended
  Consequences of the {GDPR}.}
\newblock \emph{Management Science} 69~(10): 5695--5721.
\newblock \url{https://doi.org/10.1287/mnsc.2023.4709}.

\bibitem[{Jones and Tonetti(2020)}]{Jones20}
Jones, Charles~I. and Christopher Tonetti. 2020.
\newblock \enquote{Nonrivalry and the Economics of Data.}
\newblock \emph{American Economic Review} 110~(9): 2819--2858.
\newblock \url{https://doi.org/10.1257/aer.20191330}.

\bibitem[{Kr\"{a}mer and Shekhar(2025)}]{Kramer25}
Kr\"{a}mer, Jan and Shiva Shekhar. 2025.
\newblock \enquote{Regulating Digital Platform Ecosystems Through Data Sharing
  and Data Siloing: Consequences for Innovation and Welfare.}
\newblock \emph{MIS Quarterly} 49~(1): 123--154.
\newblock \url{https://doi.org/10.25300/MISQ/2024/18428}.

\bibitem[{Ma and Wu(2025)}]{Ma25}
Ma, Guang and Hong Wu. 2025.
\newblock \enquote{Cross-Border Data Flow Supervision in China's Free Trade
  Zones: Security and Compliance Rules.}
\newblock \emph{Asia Pacific Law Review} 33~(2): 231--262.
\newblock \url{https://doi.org/10.1080/10192557.2025.2471312}.

\bibitem[{Mattoo and Meltzer(2018)}]{Mattoo18}
Mattoo, Aaditya and Joshua~P. Meltzer. 2018.
\newblock \enquote{International Data Flows and Privacy: The Conflict and Its
  Resolution.}
\newblock \emph{Journal of International Economic Law} 21~(4): 769--789.
\newblock \url{https://doi.org/10.1093/jiel/jgy044}.

\bibitem[{Ngai et~al.(2011)Ngai, Hu, Wong, Chen, and Sun}]{Ngai11}
Ngai, E.W.T., Yong Hu, Y.H. Wong, Yijun Chen, and Xin Sun. 2011.
\newblock \enquote{The Application of Data Mining Techniques in Financial Fraud
  Detection: A Classification Framework and an Academic Review of Literature.}
\newblock \emph{Decision Support Systems} 50~(3): 559--569.
\newblock \url{https://doi.org/10.1016/j.dss.2010.08.006}.

\bibitem[{{OECD}(2023)}]{OECD23}
{OECD}. 2023.
\newblock \enquote{The Nature, Evolution and Potential Implications of Data
  Localisation Measures.}
\newblock , OECD Publishing, Paris.
\newblock \url{https://doi.org/10.1787/179f718a-en}.

\bibitem[{{OECD/WTO}(2025)}]{OECDWTO25}
{OECD/WTO}. 2025.
\newblock \enquote{Economic Implications of Data Regulation: Balancing Openness
  and Trust.}
\newblock , OECD Publishing and World Trade Organization, Paris and Geneva.
\newblock \url{https://doi.org/10.1787/aa285504-en}.

\bibitem[{Prince and Wallsten(2025)}]{Prince25}
Prince, Jeffrey~T. and Scott Wallsten. 2025.
\newblock \enquote{Do People Around the World Care Where Their Data Are
  Stored?}
\newblock \emph{Information Economics and Policy} 71: 101132.
\newblock \url{https://doi.org/10.1016/j.infoecopol.2025.101132}.

\bibitem[{Rong et~al.(2025)Rong, Ling, Yang, and Huang}]{Rong25}
Rong, Ke, Yunshu Ling, Tianxi Yang, and Cheng Huang. 2025.
\newblock \enquote{Cross-Border Data Transfer: Patterns and Discrepancies.}
\newblock \emph{Journal of International Business Policy} 8~(1): 10--32.
\newblock \url{https://doi.org/10.1057/s42214-025-00209-7}.

\bibitem[{Rudin(2019)}]{Rudin19}
Rudin, Cynthia. 2019.
\newblock \enquote{Stop Explaining Black Box Machine Learning Models for High
  Stakes Decisions and Use Interpretable Models Instead.}
\newblock \emph{Nature Machine Intelligence} 1~(5): 206--215.
\newblock \url{https://doi.org/10.1038/s42256-019-0048-x}.

\bibitem[{Siering(2022)}]{Siering22}
Siering, Michael. 2022.
\newblock \enquote{Explainability and Fairness of {RegTech} for Regulatory
  Enforcement: Automated Monitoring of Consumer Complaints.}
\newblock \emph{Decision Support Systems} 158: 113782.
\newblock \url{https://doi.org/10.1016/j.dss.2022.113782}.

\bibitem[{{Standing Committee of the National People's Congress
  (NPC)}(2021)}]{NPC21}
{Standing Committee of the National People's Congress (NPC)}. 2021.
\newblock \enquote{Personal Information Protection Law of the People's Republic
  of China.}
\newblock , Standing Committee of the National People's Congress, Beijing.
\newblock
  \url{http://www.npc.gov.cn/npc/c2/c30834/202108/t20210820_313088.html}.
\newblock Adopted 20 August 2021, effective 1 November 2021.

\bibitem[{Stolz et~al.(2024)Stolz, Krasowski, Thumm, Eichelbeck, Gassert, and
  Althoff}]{Stolz24}
Stolz, Roland, Hanna Krasowski, Jakob Thumm, Michael Eichelbeck, Philipp
  Gassert, and Matthias Althoff. 2024.
\newblock \enquote{Excluding the Irrelevant: Focusing Reinforcement Learning
  Through Continuous Action Masking.}
\newblock In \emph{Advances in Neural Information Processing Systems 37
  (NeurIPS 2024)}, pp. 95067--95094. Red Hook, NY: Curran Associates.
\newblock \url{https://doi.org/10.52202/079017-3013}.

\bibitem[{Wachi, Shen, and Sui(2024)}]{Wachi24}
Wachi, Akifumi, Xun Shen, and Yanan Sui. 2024.
\newblock \enquote{A Survey of Constraint Formulations in Safe Reinforcement
  Learning.}
\newblock In \emph{Proceedings of the 33rd International Joint Conference on
  Artificial Intelligence (IJCAI-24), Survey Track}, pp. 8262--8271.
\newblock \url{https://doi.org/10.24963/ijcai.2024/913}.

\bibitem[{Zhou, Fotouhi, and Miller-Hooks(2025)}]{Zhou25}
Zhou, Weiwen, Hossein Fotouhi, and Elise Miller-Hooks. 2025.
\newblock \enquote{Decision Support Through Deep Reinforcement Learning for
  Maximizing a Courier's Monetary Gain in a Meal Delivery Environment.}
\newblock \emph{Decision Support Systems} 190: 114388.
\newblock \url{https://doi.org/10.1016/j.dss.2024.114388}.

\end{thebibliography}
\appendix

\setcounter{algorithm}{0}
\renewcommand{\thealgorithm}{A-\arabic{algorithm}}

\section{Algorithms of the solution method}\label{a:appendixA}

The solution method comprises two modules (Fig.~\ref{f:fig2}). The counterfactual path-prediction module computes the long-run values of the four strategic paths offline and trains a predictor on them (Algorithm~\ref{alg:A1}); the masked D3QN+CPAA agent reads the predictions online and decides week by week within the legal action set (Algorithm~\ref{alg:A2}).

\begin{algorithm}[H]
\caption{Offline label generation and predictor training for the counterfactual path advantages}
\label{alg:A1}
\begin{algorithmic}[1]
\Require Default continuation policy $\pi_0$; label discount $\gamma=0.99$; horizon $T=52$; training firms $\mathcal{D}$
\Ensure Counterfactual path-advantage predictor $f_\psi$
\State Initialize label set $\mathcal{L} \leftarrow \emptyset$
\For{each firm in $\mathcal{D}$}
    \State Sample the firm's $T$-week exogenous task stream
    \For{$t = 0, 1, \ldots, T-1$}
        \State Observe base state $\tilde s_t$; via $\mathcal{R}_{\min}$ get minimum compliance stringency $r_t$ and legal set $\mathcal{A}_{\mathrm{legal}}(s_t)$
        \State $G_0 \leftarrow G(s_t, \pi_0(s_t))$ \Comment{discounted reward following $\pi_0$ from week $t$}
        \For{path $k$ in $\{\mathrm{LOCAL}, \mathrm{L0}, \mathrm{L1}, \mathrm{L2}\}$}
            \If{representative action $a^k \in \mathcal{A}_{\mathrm{legal}}(s_t)$}
                \State $G_k \leftarrow G(s_t, a^k)$ \Comment{take $a^k$ at week $t$, then follow $\pi_0$}
                \State $\Lambda_t[k] \leftarrow G_k - G_0$; \quad $\mathrm{mask}_t[k] \leftarrow 1$
            \Else
                \State $\Lambda_t[k] \leftarrow 0$; \quad $\mathrm{mask}_t[k] \leftarrow 0$
            \EndIf
        \EndFor
        \State Build 32-dimensional feature $\phi_t$; \quad $\mathcal{L} \leftarrow \mathcal{L} \cup \{(\phi_t, \Lambda_t, \mathrm{mask}_t)\}$
    \EndFor
\EndFor
\State Train $f_\psi$ on $\mathcal{L}$ by masked MSE; clip predictions to $[-5, 5]$
\State \Return $f_\psi$
\end{algorithmic}
\end{algorithm}

\begin{algorithm}[H]
\caption{Training of the masked D3QN+CPAA}
\label{alg:A2}
\begin{algorithmic}[1]
\Require Frozen predictor $f_\psi$; episodes $E=3000$; horizon $T=52$; discount $\gamma=0.99$
\Ensure Policy value network $Q_\theta$
\State Initialize online net $Q_\theta$, target net $Q_{\theta^-} \leftarrow Q_\theta$, replay buffer $\mathcal{B} \leftarrow \emptyset$; $\epsilon \leftarrow 0.2$
\For{episode $= 1, \ldots, E$}
    \State Reset environment: $Q^{pi}, Q^{spi} \leftarrow 0$; draw first-week task
    \For{$t = 0, 1, \ldots, T-1$}
        \State Observe $s_t = (\tilde s_t, r_t)$; $\hat\Lambda_t \leftarrow \operatorname{clip}(f_\psi(\phi_t), -5, 5)$; $z_t \leftarrow (s_t, \hat\Lambda_t)$
        \State Via $\mathcal{R}_{\min}$ get legal set $\mathcal{A}_{\mathrm{legal}}(s_t)$
        \If{with probability $\epsilon$}
            \State Pick $a_t$ uniformly from $\mathcal{A}_{\mathrm{legal}}(s_t)$
        \Else
            \State $a_t \leftarrow \arg\max_{a \in \mathcal{A}_{\mathrm{legal}}(s_t)} Q_\theta(z_t, a)$
        \EndIf
        \State Execute $a_t$; receive reward $R_t$ and next augmented observation $z_{t+1}$
        \State $\mathcal{B} \leftarrow \mathcal{B} \cup \{(z_t, a_t, R_t, z_{t+1}, \mathcal{A}_{\mathrm{legal}}(s_{t+1}))\}$
        \State Sample a minibatch from $\mathcal{B}$; compute double, masked bootstrap target $y_t$
        \State Update $\theta$ by the gradient of $(Q_\theta(z_t, a_t) - y_t)^2$
        \State Every 1000 steps synchronize $Q_{\theta^-} \leftarrow Q_\theta$
    \EndFor
    \State $\epsilon \leftarrow \mathrm{linear\_anneal}(0.2 \to 0.05)$
\EndFor
\State \Return $Q_\theta$
\end{algorithmic}
\end{algorithm}

\section{Network architectures, action discretization, and training setup}\label{a:appendixB}

\setcounter{table}{0}
\renewcommand{\thetable}{B-\arabic{table}}

This appendix provides the implementation details of the solution method, including the environment calibration parameters, the action discretization, and the network architectures and hyperparameters of the counterfactual path-prediction module and the masked RL learners. Augmented (+CPAA) and unaugmented configurations share the same hyperparameters except for the observation dimension. Every learner is trained once under each of five randomly chosen seeds, with each run budgeted at 3000 episodes (52 weeks per episode, i.e., $1.56\times10^{5}$ environment steps), and evaluated deterministically on the 300 held-out test firms. Table~\ref{t:tableB1} lists the environment calibration parameters used in the numerical experiments.

\begin{table}[H]
\caption{Environment calibration parameters}
\begin{tabular*}{\textwidth}[]{@{\extracolsep\fill}lcc}
\toprule
Parameter & Symbol & Value\\
\midrule
Reference scale & $q_{\mathrm{ref}}$ & $5\times10^{4}$\\
Horizon & $T$ & $52$ weeks\\
Friction memory & $\alpha$ & $0.85$\\
Value curvature & $\beta$ & $3.0$\\
Shortfall rate & $\mu$ & $0.30$\\
Persistent-friction weight & $\kappa_A$ & $0.50$\\
Per-period friction weight & $\kappa_\sigma$ & $0.30$\\
Credential invalidation rate & $p_{\mathrm{chg}}$ & $0.08$\\
Level-1 acquisition cost & $F(1)$ & $0.35$\\
Level-2 acquisition cost & $F(2)$ & $0.55$\\
\bottomrule
\end{tabular*}
\label{t:tableB1}\end{table}

The transfer volume $x_t$ is discretized into 10 equally spaced levels on $[0,q_t]$; composed with the five response paths, this yields $5\times10=50$ discrete mechanism--volume actions (local processing fixes $x_t=0$), on which all masked RL learners operate. Writing the observation dimension as $d_{\mathrm{in}}$, the unaugmented configurations take the 13-dimensional base state $s_t$ as input, and the +CPAA configurations append the 4-dimensional counterfactual path advantages $\hat\Lambda_t$ for 17 dimensions in total; the output layer scores the 50 mechanism--volume actions, the value networks and critic networks output a single value, and all hidden layers use ReLU activations. The legal mask (Eq.~\eqref{e:eq10}) is applied state by state before every action selection, screening out export mechanisms with strength below $r_t$, so illegal exports remain unreachable throughout training and evaluation.

The predictor of the counterfactual path-prediction module is a three-hidden-layer multilayer perceptron with hidden widths 256, 128, and 64, each followed by a ReLU activation, taking the 32-dimensional feature $\phi_t$ as input and outputting the counterfactual advantages of the four paths. It is trained by masked mean squared error for 20 epochs with learning rate $10^{-3}$, batch size 256, and label discount factor $\gamma=0.99$; predictions are truncated to $[-5,5]$. All labels are computed once offline and cached, over 3000 training firms $\times$ 52 weeks for 156000 samples in total.

Table~\ref{t:tableB2} lists the network architectures and core hyperparameters of the six masked RL learners. For fair comparison, all methods use the same learning rate $5\times10^{-4}$ and discount factor $0.99$. The on-policy methods (PPO, A2C) update on 52-step rollouts without experience replay; PPO uses a clipping coefficient of $0.2$, GAE $\lambda=0.95$, an entropy coefficient of $0.01$, and 10 epochs per update. The value-based methods (DQN family) synchronize the target network every 1000 steps, warm up the replay buffer for 1000 steps, and anneal the $\epsilon$-greedy exploration rate linearly from $0.2$ to $0.05$.

\begin{table}[H]
\caption{Network architectures and hyperparameters of the masked RL learners}
\begin{tabular*}{\textwidth}[]{@{\extracolsep\fill}lp{4.6cm}cccc}
\toprule
Method & Architecture & Learning rate & Discount & Replay capacity & Batch size\\
\midrule
DQN & $d_{\mathrm{in}}\times256\times256\times50$ & $5\times10^{-4}$ & $0.99$ & $50000$ & $64$\\
Double DQN & $d_{\mathrm{in}}\times256\times256\times50$ & $5\times10^{-4}$ & $0.99$ & $50000$ & $64$\\
Dueling DQN & $d_{\mathrm{in}}\times256\times256$, value head $\to1$, advantage head $\to50$ & $5\times10^{-4}$ & $0.99$ & $50000$ & $64$\\
D3QN & $d_{\mathrm{in}}\times256\times256$, value head $\to1$, advantage head $\to50$ & $5\times10^{-4}$ & $0.99$ & $50000$ & $64$\\
PPO & actor $d_{\mathrm{in}}\times128\times128\times50$, critic $d_{\mathrm{in}}\times64\times64\times1$ & $5\times10^{-4}$ & $0.99$ & N/A & $52$\\
A2C & actor $d_{\mathrm{in}}\times128\times128\times50$, critic $d_{\mathrm{in}}\times64\times64\times1$ & $5\times10^{-4}$ & $0.99$ & N/A & $52$\\
\bottomrule
\end{tabular*}
\label{t:tableB2}\end{table}

\end{document}